\documentclass[usenatbib,useAMS,letterpaper]{mnras}
\usepackage{float}
\usepackage{graphicx}
\usepackage{epstopdf}
\usepackage{ textcomp }
\usepackage{multirow}
\usepackage{ctable}
\usepackage{url}
\usepackage{times}
\usepackage{amsmath}
\usepackage{amssymb}
\usepackage{xspace}
\usepackage{mathrsfs} 
\usepackage{tabularx}
\usepackage{fixltx2e}
\usepackage{color}
\usepackage{placeins}
\usepackage{comment}

\newcommand{\acknowledgments}{\begin{small}\section*{Acknowledgements}\end{small}}

\defcitealias{2019MNRAS.487.4393S}{Paper {\small I}}
\defcitealias{2020MNRAS.491.1190S}{Paper {\small II}}
\defcitealias{2021MNRAS.507..175S}{Paper {\small III}}
\newcommand\sref[1]{\hyperref[#1]{\S~\ref*{#1}}}
\newcommand\fref[1]{\hyperref[#1]{Fig.~\ref*{#1}}}
\newcommand\Eqref[1]{Eq.~(\hyperref[#1]{\ref*{#1}})}
\newcommand\eeqref[1]{Eq.~\hyperref[#1]{\ref*{#1}}}

\newcommand\tref[1]{\hyperref[#1]{Table~\ref*{#1}}}
\newcommand\aref[1]{\hyperref[#1]{Appendix~\ref*{#1}}}

\newcommand{\hlt}[1]{\textcolor{blue}{#1}}

\newcommand{\oneline}[1]{%
  \newdimen{\namewidth}%
  \setlength{\namewidth}{\widthof{#1}}%
  \ifthenelse{\lengthtest{\namewidth < \textwidth}}%
  {#1}
  {\resizebox{\textwidth}{!}{#1}}
}
 
\title[What Types of Jet Quench?]{Unraveling Jet Quenching Criteria Across L* Galaxies and Massive Cluster Ellipticals}
 
\author[]{
\parbox[t]{\textwidth}{
Kung-Yi Su$^{1}$\thanks{E-mail: kungyisu@g.harvard.edu}, Greg L. Bryan$^2$, Christopher C. Hayward$^3$, Rachel S. Somerville$^3$, Philip F. Hopkins$^4$, Razieh \ Emami$^5$, Claude-Andr\'{e} Faucher-Gigu\`{e}re$^6$, Eliot Quataert$^7$, Sam B. Ponnada$^4$, Drummond Fielding$^3$, Du\v{s}an Kere\v{s}$^8$
}
\vspace*{6pt} \\
$^1$Black Hole Initiative, Harvard University, 20 Garden Street, Cambridge, MA 02138, USA\\
$^2$Department of Astronomy, Columbia University, 550 West 120th Street, New York, NY 10027, USA\\
$^3$Center for Computational Astrophysics, Flatiron Institute, 162 Fifth Avenue, New York, NY 10010, USA\\
$^4$TAPIR 350-17, California Institute of Technology, 1200 E. California Boulevard, Pasadena, CA 91125, USA\\
$^5$ Center for Astrophysics $\vert$ Harvard \& Smithsonian, 60 Garden Street, Cambridge, MA 02138, USA\\
$^6$Department of Physics \& Astronomy and CIERA, Northwestern University, 1800 Sherman Ave, Evanston, IL 60201, USA\\
$^7$ Department of Astrophysical Sciences, Princeton University, Princeton, NJ 08544, USA\\
$^8$Department of Physics and Center for Astrophysics and Space Science, University of California at San Diego, 9500 Gilman Drive, La Jolla, CA 92093, USA
}
\usepackage[all]{hypcap}

\begin{document}
\long\def\/*#1*/{}
\date{Submitted to MNRAS}

\pagerange{\pageref{firstpage}--\pageref{lastpage}} \pubyear{2021}

\maketitle

\label{firstpage}

\begin{abstract}
In the absence of supplementary heat, the radiative cooling of halo gas around massive galaxies (Milky Way mass and above) leads to an excess of cold gas or stars beyond observed levels. AGN jet-induced heating is likely essential, but the specific properties of the jets remain unclear. Our previous work \citep{2021MNRAS.507..175S} concludes from simulations of a halo with $10^{14} M_\odot$ that a successful jet model should have an energy flux comparable to the free-fall energy flux at the cooling radius and should inflate a sufficiently wide cocoon with a long enough cooling time. In this paper, we investigate three jet modes with constant fluxes satisfying the criteria, including high-temperature thermal jets, cosmic ray (CR)-dominant jets, and widely precessing kinetic jets in  $10^{12}-10^{15}\,{\rm M}_{\odot}$ halos using high-resolution, non-cosmological MHD simulations with the FIRE-2 (Feedback In Realistic Environments) stellar feedback model, conduction, and viscosity. 
We find that scaling the jet energy according to the free-fall energy at the cooling radius can successfully suppress the cooling flows and quench galaxies without obviously violating observational constraints. 
We investigate an alternative scaling method in which we adjust the energy flux based on the total cooling rate within the cooling radius. However, we observe that the strong interstellar medium (ISM) cooling dominates the total cooling rate in this scaling approach, resulting in a jet flux that exceeds the amount needed to suppress the cooling flows. With the same energy flux, the CR-dominant jet is most effective in suppressing the cooling flow across all the surveyed halo masses due to the enhanced CR pressure support. 
We confirm that the criteria for a successful jet model, which we proposed in \cite{2021MNRAS.507..175S}, work across a much wider range, encompassing halo masses of $10^{12}-10^{15} {\rm M_\odot}$.

\end{abstract}


\section{Introduction}
\label{S:intro}
For years, the perplexing issue in galaxy formation has revolved around effectively ``quenching'' massive galaxies (with stellar masses $\gtrsim 10^{11}\,{\rm M}{\odot}$ or exceeding $\sim L{\ast}$ in the galaxy luminosity function) and maintaining them in a ``red and dead'' state across a large fraction of cosmic time \citep[see, e.g.,][]{2003ApJS..149..289B,2003MNRAS.341...54K,2003MNRAS.343..871M,2004ApJ...600..681B,2005MNRAS.363....2K,2005ApJ...629..143B,2006MNRAS.368....2D,2009MNRAS.396.2332K,2010A&A...523A..13P,2012MNRAS.424..232W,2015MNRAS.446.1939F,2015Natur.519..203V}. This challenge stems from the persistent ``cooling flow'' problem, as evidenced by X-ray observations revealing substantial radiative cooling in the hot circum-galactic medium (CGM) or intra-cluster medium (ICM) gas of elliptical galaxies and clusters, indicating cooling times shorter than a Hubble time \citep{1994ApJ...436L..63F,2006PhR...427....1P,2019MNRAS.488.2549S}. Despite the inferred cooling flow, with rates reaching up to $\sim 1000\,{\rm M}_\odot {\rm yr}^{-1}$ in clusters, observations of galaxies show a shortage of both cold gas from H{\scriptsize I} and CO \citep{2011ApJ...731...33M,2013ApJ...767..153W}, and a lack of significant star formation \citep{2001A&A...365L..87T,2008ApJ...681.1035O,2008ApJ...687..899R}. Notably, simulations and semi-analytic models that neglect to suppress cooling flows and permit gas to freely cool into the galactic core consistently predict star formation rates (SFRs) over an order of magnitude higher than observed \citep[see recent examples, such as the weak/no feedback runs in][]{2007MNRAS.380..877S, somerville:2008, 2009MNRAS.398...53B, 2015MNRAS.449.4105C, 2015ApJ...811...73L, 2017MNRAS.472L.109A}.

To counteract the observed cooling, it is imperative to have a heat source or pressure support. Additionally, the heating process must maintain the inherent structure of a cool core, such as density and entropy profiles, as observed in the majority of galaxies (e.g., \citep{1998MNRAS.298..416P,2009A&A...501..835M}). Various non-AGN feedback mechanisms proposed in existing literature, such as stellar feedback from shock-heated AGB winds, Type Ia supernovae (SNe), SNe-injected cosmic rays (CRs), magnetic fields, and thermal conduction in the CGM or ICM, or the concept of ``morphological quenching,'' have proven ineffective in addressing the cooling flow problem (\citealt{2019MNRAS.487.4393S}, hereafter referred to as \citetalias{2019MNRAS.487.4393S}). 
Consequently, AGN feedback emerges as the most promising solution, supported by an extensive body of theoretical work (for recent studies, refer to subsequent paragraphs on AGN jet and, e.g., \citealt[][]{2017ApJ...837..149G,2017MNRAS.468..751E,2018MNRAS.479.4056W,2018ApJ...866...70L,2018ApJ...856..115P,2018ApJ...864....6Y} for other forms of AGN feedback; also see, e.g., \citealt{1998A&A...331L...1S,1999MNRAS.308L..39F,2001ApJ...551..131C,2005ApJ...630..705H,2006ApJS..163....1H,2006MNRAS.365...11C,2009ApJ...699...89C,2012ApJ...754..125C} for earlier works). 

Observational studies further indicate that the energy budget from AGN aligns with the cooling rate \citep{2004ApJ...607..800B}. Additionally, there are documented instances of AGN expelling gas from galaxies, introducing thermal energy through shocks or sound waves, employing photo-ionization and Compton heating, or influencing the CGM and ICM through stirring, creating hot plasma ``bubbles'' with substantial relativistic components?phenomena consistently observed around massive galaxies \citep[see, e.g.,][for a comprehensive review]{2012ARA&A..50..455F,2018ARA&A..56..625H}.

However, despite its plausibility and the extensive work conducted previously, the detailed physics of AGN feedback remains uncertain, as do the relevant ``input parameters.'' In our subsequent works \citetalias{2020MNRAS.491.1190S} \citep{2020MNRAS.491.1190S} and \citetalias{2021MNRAS.507..175S} \citep{2021MNRAS.507..175S}, we tested a wide variety of AGN feedback models. In particular, in \citetalias{2021MNRAS.507..175S}, we focused our study on testing a wide variety of jet models with a set of isolated galaxy simulations of massive cluster ellipticals with a halo mass of $10^{14}M_\odot$. We conducted tests on various constant-energy-flux jet models, systematically altering factors such as the jet energy form (kinetic, thermal, cosmic ray), energy, momentum, and mass fluxes, magnetic field strength and geometry, opening angle, precession, and duty cycle. While certain aspects of jets have been previously investigated independently \citep[e.g.,][]{2012ApJ...746...94G,2014ApJ...789...54L,2016ApJ...818..181Y,2017MNRAS.472.4707B,2017ApJ...844...13R,2019MNRAS.483.2465M}, our set of simulations stands out as the first to explore all variations of jet models while concurrently incorporating explicit stellar feedback (FIRE; \citealt{2017arXiv170206148H}) and fluid microphysics. In that work, we found that the criteria for a successful jet model are: (i) an optimal energy flux comparable to the free fall energy flux at the cooling radius and (ii) a sufficiently wide jet cocoon with a long enough cooling time at the cooling radius (summarized in \sref{s:criteria}). To fulfill these criteria, we found that we needed either a CR-dominant jet, a very hot thermal energy-dominant jet, or a widely precessing jet (\citetalias{2021MNRAS.507..175S}). Among these possibilities, CR dominant jets can most easily fulfill the above criteria allowing a wider parameter space because of the extra pressure and the suppression of thermal instability.

A major missing piece of \citetalias{2021MNRAS.507..175S} is the relatively narrow halo mass range explored ($10^{14} M_\odot$), which limited the applicability of our conclusion in not only galaxy type, but also evolutionary stage of the galaxy. The halo mass could affect the cooling flow properties in several aspects. At a halo mass of $\sim10^{12} M_\odot$, where the cooling flow problem begins, although the net CGM mass is not as high as the more massive systems, the virial temperature sits near the peak of the cooling curve, resulting in a shorter cooling time. As the halo mass increases, the gas available in the CGM increases, but the cooling curve drops until the halo mass is around $10^{13}-10^{14} M_\odot$, beyond which the cooling rate increases again. As a result, the cooling flow properties of the gas halos with $\sim10^{12}-10^{15} M_\odot$ can differ significantly in their cooling flow properties, and a systematic study would be useful. Moreover, there are more constraints from X-ray observations for rich clusters of mass $\sim 10^{15}{\rm M}_\odot$. We, therefore, study our various ``more successful'' jet models with a broad range of halo masses ($10^{12}-10^{15}\,{\rm M}_{\odot}$) and test how they should be scaled to stably suppress the cooling flow and maintain the galaxy as quenched in this study.

In \sref{S:methods}, we summarize our initial conditions (ICs) and the AGN jet parameters we survey and describe our numerical simulations.  We present our results and describe the observational properties of the runs in \sref{S:results}. We discuss the quenching criteria in \sref{s:criteria_mass}. We explore the limitations of this work and potential future research in \sref{s:discussion}. Finally, we conclude in \sref{sec:conclusions}

\section{Methodology} \label{S:methods}
We perform simulations on isolated galaxies, each possessing a halo mass ranging from approximately $10^{12}$ to $10^{15} {\rm M}_\odot$. The initial conditions are configured based on the observed profiles of cool-core clusters at low redshift, as outlined in \sref{S:ic}. In the absence of AGN feedback, despite the galaxies exhibiting initial properties consistent with observations, their cooling flow rates and SFRs rapidly escalate, surpassing the observational values of quenched populations by orders of magnitude (\citetalias{2019MNRAS.487.4393S} and \citetalias{2020MNRAS.491.1190S}). We progress the simulations using various AGN jet models, aiming to assess the extent to which they inhibit the cooling flow and whether they can sustain stably quenched galaxies

Our simulations utilize {\sc GIZMO}\footnote{A publicly accessible version of this code can be found at \href{http://www.tapir.caltech.edu/~phopkins/Site/GIZMO.html}{\textit{http://www.tapir.caltech.edu/$\sim$phopkins/Site/GIZMO.html}}} \citep{2015MNRAS.450...53H} in its meshless finite mass (MFM) mode. This mode employs a Lagrangian mesh-free Godunov method, capturing the benefits of both grid-based and smoothed-particle hydrodynamics (SPH) methods. Details regarding numerical implementation and thorough testing are presented in a series of method papers covering various aspects such as hydrodynamics and self-gravity \citep{2015MNRAS.450...53H}, magnetohydrodynamics \citep[MHD;][]{2016MNRAS.455...51H,2015arXiv150907877H}, anisotropic conduction and viscosity \citep{2017MNRAS.466.3387H,2017MNRAS.471..144S}, and cosmic rays \citep{chan:2018.cosmicray.fire.gammaray}.

In summary, for cosmic rays, we track a single energy bin of GeV cosmic ray protons, which exert dominant pressure, and treat them within the ultra-relativistic limit. The cosmic ray transport involves streaming at the local Alfv'en speed, incorporating the appropriate streaming loss term, thermalizing according to \citet{Uhlig2012}, but with $v_{\rm st}=v_A$. Additionally, the model accounts for diffusion with a fixed diffusivity denoted as $\kappa_{\rm CR}$, adiabatic energy exchange with both gas and cosmic ray pressure, and hadronic and Coulomb losses following \citet{2008MNRAS.384..251G}. Both streaming and diffusion occur fully anisotropically along magnetic field lines. In previous works such as \citet{chan:2018.cosmicray.fire.gammaray, hopkins:cr.mhd.fire2, 2021MNRAS.501.4184H}, we demonstrated that achieving observed $\gamma$-ray luminosities in simulations with the outlined physics necessitates $\kappa_{\rm CR}\sim 10^{29}\,{\rm cm^{2}\,s^{-1}}$. This value aligns well with detailed cosmic ray transport models that encompass an extended gaseous halo around the Galaxy, as evidenced in \citep[see, for example,][]{1998ApJ...509..212S, 2010ApJ...722L..58S, 2011ApJ...729..106T}. Consequently, we adopt this as our fiducial value and discuss uncertainties in \sref{s:caveat}.

All our simulations incorporate the FIRE-2 implementation of the Feedback In Realistic Environments (FIRE) physical treatments for the interstellar medium (ISM), star formation, and stellar feedback. The comprehensive details of these treatments can be found in \citet{hopkins:sne.methods,2017arXiv170206148H}, accompanied by extensive numerical tests. Cooling is tracked across the temperature range of $10-10^{10}$ K, accounting for various processes such as photo-electric and photo-ionization heating, collisional, Compton, fine structure, recombination, atomic, and molecular cooling. Star formation is implemented through a sink particle method, allowed only in molecular, self-shielding, locally self-gravitating gas with a density greater than $100\,{\rm cm^{-3}}$ \citep{2013MNRAS.432.2647H}. Once formed, star particles are treated as a single stellar population with metallicity inherited from their parent gas particle at formation. All feedback rates, including supernovae (SNe) and mass-loss rates, as well as spectra, are IMF-averaged values calculated from {\small STARBURST99} \citep{1999ApJS..123....3L} with a \citet{2002Sci...295...82K} initial mass function (IMF). The stellar feedback model encompasses (1) Radiative feedback, encompassing photo-ionization and photo-electric heating, along with tracking single and multiple-scattering radiation pressure in five bands (ionizing, FUV, NUV, optical-NIR, IR), (2) OB and AGB winds, leading to continuous stellar mass loss and injection of mass, metals, energy, and momentum, and (3) Type II and Type Ia supernovae (including both prompt and delayed populations) occurring based on tabulated rates, injecting the appropriate mass, metals, momentum, and energy into the surrounding gas.

\subsection{Initial Conditions}
\label{S:ic}

\begin{figure}\label{fig:xray}
    \centering
    \includegraphics[width=8cm]{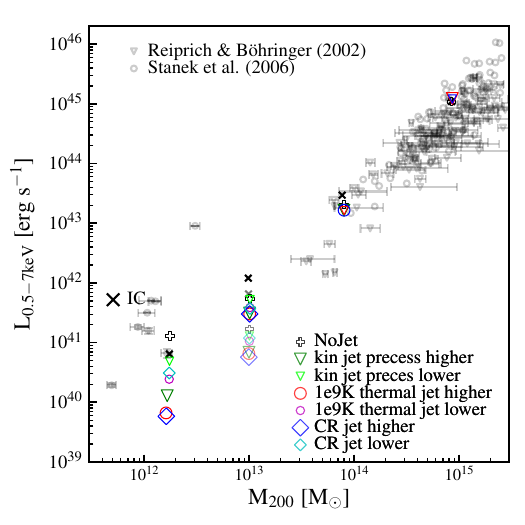}
    \caption{The X-ray luminosity in the 0.5 - 7 keV band at the end of all the non-overheated quiescent runs. We use $M_{\rm 200}$ as the halo mass for our simulations. The lighter markers and the error bars denote the observed values from \citep{2002ApJ...567..716R,2006ApJ...648..956S}. m13l run is colored with light color than m13. The X-ray luminosities of the initial conditions are within the observational range.
    We observe very little evolution of the total X-ray luminosity for most runs except for m13l and m12 runs with higher jet energy fluxes (compared to the values for the initial conditions), where the resulting X-ray luminosities are slightly lower than the observational range. }
    \label{fig:res_acc}
\end{figure}

The initial conditions we explore are thoroughly detailed in \citetalias{2019MNRAS.487.4393S} and \citetalias{2020MNRAS.491.1190S}. We made slight adjustments to the initial conditions employed in the previous two papers by extending the galaxy model to a larger radius and increasing the box size, following the approach used for the `m14' run in \citetalias{2021MNRAS.507..175S}. These initial conditions aim to closely resemble observed cool-core systems of comparable mass whenever possible at $z\sim0$ \citep[see, e.g.,][]{2012ApJ...748...11H,2013MNRAS.436.2879H,2013ApJ...775...89S,2015ApJ...805..104S,2017A&A...603A..80M}. A summary of their properties is provided in \tref{tab:ic}. The initialization of the dark matter (DM) halo, bulge, black hole, and gas+stellar disk follows the method outlined by \cite{1999MNRAS.307..162S} and \cite{2000MNRAS.312..859S} with live particles.

We assume a spherical NFW \citep{1996ApJ...462..563N} profile for the DM halo, a \cite{1990ApJ...356..359H} profile for the stellar bulge, and an exponential, rotation-supported disk of gas and stars initialized with Toomre $Q\approx1$. The black hole (BH) is assigned a mass approximately $1/300$ of the stellar mass \citep[e.g.,][]{2004ApJ...604L..89H}. The extended spherical, hydrostatic gas halo is modeled with a $\beta$-profile and rotation at twice the net DM spin, ensuring that below 10-15\% of the support against gravity comes from rotation, with the majority of support originating from thermal pressure, as expected in a massive halo. Additionally, we introduce a ``m13-mBH'' (m13 massive BH) model, artificially increasing the black hole mass to $5\times10^9 M_\odot$ to investigate the impact of black hole mass (while maintaining a fixed jet mass flux for ease of comparison). The parameters mentioned above are summarized in \tref{tab:ic}.

The initial metallicity of the CGM/ICM decreases from solar ($Z=0.02$) to $Z=0.001$ with radius, following the expression $Z(r)=0.02\,(0.05+0.95/(1+(r/r_Z\,{\rm })^{1.5}))$, where $r_z=(20,10,10,20)$ kpc for (m12, m13l, m13, m14). For m15, the initial metallicity profile is defined as $Z(r)=0.02\,(0.05+0.5/(1+(r/r_{Z0}\,{\rm })^{1.5})+0.45/(1+(r/r_{Z1}\,{\rm })^{1.5}))$, with $r_{Z0}=10$ kpc and $r_{Z1}=750$ kpc.

The initial magnetic fields are azimuthal with a seed value of $|{\bf B}|=B_0/(1+(r/r_B)^{2})^{\beta_B}$ (to be later amplified) extending throughout the ICM, where $B_0=(0.03, 0.1, 0.1, 0.3) {\rm \mu G}$, $r_B=20 {\rm kpc}$, and $\beta_B=(0.375, 1,1,0.375)$ for (m12, m13l, m13, m14). The field strength roughly scales as $B_0\propto M_{\rm halo}^{0.5}$. For m15, the initial magnetic field strength is set to achieve a plasma $\beta=1000$.

The initial cosmic ray energy density is in equipartition with the local initial magnetic energy density. Both the seed magnetic energy and cosmic ray energy are weak compared to the thermal energy. The initial conditions are evolved adiabatically (no cooling or star formation) for 50 Myr to allow for the relaxation of any initial transients.

 As shown in \fref{fig:xray}, our initial conditions all have an initial  X-ray luminosity in 0.5-7 keV band consistent with the observations.

A resolution analysis is appended in \citetalias{2019MNRAS.487.4393S}. For m12, m13l, and m13, a fixed mass resolution of $(8\times10^3$, $5\times10^4$, and $5\times10^4)$ $M_\odot$ is employed. In the case of m14 and m15, a hierarchical super-Lagrangian refinement scheme (\citetalias{2019MNRAS.487.4393S} and \citetalias{2020MNRAS.491.1190S}) is utilized to attain a mass resolution of approximately $3 \times 10^{4}\,{\rm M}{\odot}$ in the core region and around the z-axis, where the jet is launched. This resolution is significantly higher than that of many prior global studies.  The mass resolution decreases as a function of both the radius ($r_{\rm 3d}$) and the distance from the z-axis ($r_{\rm 2d}$), roughly scaling with $r_{\rm 3d}$ and $2^{r_{\rm 2d}/10 {\rm kpc}}$, whichever is smaller, reaching a minimum of $2\times 10^6 {\rm M}_{\odot}$. The region with the highest resolution corresponds to cases where either $r_{\rm 3d}$ or $r_{\rm 2d}$ is less than 10 kpc. The resolution of m15-NoJet is set at a lower level ($>5\times10^{5} M_\odot$) due to the computational cost associated with dense core gas resulting from strong cooling flows.

\begin{table*}
\begin{center}
 \caption{Properties of Initial Conditions for the Simulations/Halos Studied In This Paper}
 \label{tab:ic}
 \resizebox{17.7cm}{!}{
\begin{tabular}{lcccccccccccccccc}
 \hline
\hline
&&\multicolumn{2}{c}{\underline{Resolution}}&\multicolumn{3}{c}{\underline{DM halo}}&&\multicolumn{2}{c}{\underline{Stellar Bulge}}&\multicolumn{2}{c}{\underline{Stellar Disc}}&\multicolumn{2}{c}{\underline{Gas Disc}}&\multicolumn{3}{c}{\underline{Gas Halo}}  \\
$\,\,\,\,$Model & $R_{\rm 200}$ &$\epsilon_g^{\rm min}$ &$m_g$        &$M_{\rm DM}$   &$r_{\rm dh}$            &$\rho_0$    &$M_{\rm baryon}$    &$M_b$ 
                 &$a$          &$M_d$        & $r_d$             &$M_{\rm gd}$       &$r_{\rm gd}$         &$M_{\rm gh}$         &$r_{\rm gh}$ &$\beta$    \\
                 &(kpc)  &(pc)        &(M$_\odot$)  &(M$_\odot$)      & (kpc)           &(g/cm$^3$)           &(M$_\odot$)      &(M$_\odot$) 
                  &(kpc)        &(M$_\odot$)  &(kpc)            &(M$_\odot$)    &(kpc)            &(M$_\odot$)      &(kpc)\\
\hline
$\,\,\,\,$m12  &248       &1       &8e3           &1.5e12           &20             &5.8e-25              &2.2e11           &1.5e10   
                     &1.0       &5.0e10          &3.0                &5.0e9            &6.0                &1.5e11           &20 &0.5        \\                    
$\,\,\,\,$m13l  &444      &3       &5e4           &9.3e12           &93              &6.7e-26              &6.3e11           &1.0e11   
                     &2.8     &1.4e10        &2.8              &5.0e9           &2.8              &5.1e11              &9.3 &0.43            \\
$\,\,\,\,$m13/m13-mBH &443      &3       &5e4           &1.0e13           &87              &6.2e-26              &1.1e12           &1.0e11   
                     &2.8     &1.4e10        &2.8              &5.0e9           &2.8              &1.0e12              &9.3 &0.43            \\
{\bf $\,\,\,\,$m14} &{\bf880} &{\bf1}  &{\bf3e4}      &{\bf 6.7e13}      &{\bf 220}       &{\bf 4.3e-26}         &{\bf 9.6e12}        &{\bf2.0e11}
                     &{\bf3.9}     &{\bf2.0e10}          &{\bf3.9}              &{\bf1.0e10}           &{\bf3.9}              &{\bf9.4e12}           &{\bf 22}    &{\bf 0.5}          \\          
 $\,\,\,\,$m15  & 1955     &1       &3e4           &7.3e14           &358              &8.4e-26              &1.1e14           &2.4e11   
                     &3.14     &3.2e10        &4.9              &1.6e10           &4.9              &1.1e14              &29 & 0.56 \\                                
\hline 
\hline
\end{tabular}
}
\end{center}
\begin{flushleft}
Parameters of the galaxy/halo model studied in this paper (\sref{S:ic}): 
(1) Model name. The number following `m' labels the approximate logarithmic halo mass. 
(2) $R_{\rm 200}$. The radius that encloses an average density of 200 times the critical density.
(3) $\epsilon_g^{\rm min}$: Minimum gravitational force softening for gas (the softening for gas in all simulations is adaptive, and matched to the hydrodynamic resolution; here, we quote the minimum Plummer equivalent softening).
(4) $m_g$: Gas mass (resolution element). There is a resolution gradient for m14, so its $m_g$ is the mass of the highest resolution elements.
(5) $M_{\rm halo}$: Halo mass. 
(6) $r_{\rm dh}$: NFW halo scale radius.
(7) $V_{\rm max}$: Halo maximum circular velocity.
(8) $M_{\rm baryon}$: Total baryonic mass. 
(9) $M_b$: Bulge mass.
(10) $a$: Bulge Hernquist-profile scale-length.
(11) $M_d$ : Stellar disc mass.
(12) $r_d$ : Stellar disc exponential scale-length.
(13) $M_{\rm gd}$: Gas disc mass. 
(14) $r_{\rm gd}$: Gas disc exponential scale-length.
(15) $M_{\rm gh}$: Hydrostatic gas halo mass. 
(16) $r_{\rm gh}$: Hydrostatic gas halo $\beta=1/2$ profile scale-length.
\end{flushleft}
\end{table*}

\subsection{Implementation of AGN Jet }
\label{S:physics}
In this study, our emphasis is on examining the impact of a specific AGN jet across a broad range of halo masses. All the jet models are run with a predetermined mass, energy, and momentum flux. We do not aim to concurrently model black hole accretion from scales of $\sim10-100$ pc up to the event horizon.  

\subsubsection{Numerical implementation}
The jet is launched using a particle spawning approach, generating new gas cells (resolution elements) from the central black hole. This technique provides enhanced control over jet properties, reducing reliance on neighbor-finding outcomes. Additionally, it allows for the imposition of a higher resolution for the jet elements. The numerical methodology employed in this study bears similarity to that of \cite{2020MNRAS.497.5292T}, who investigated the impact of broad absorption line (BAL) wind feedback on disk galaxies.

The spawned gas particles exhibit a mass resolution of 5000 ${\rm M}_\odot$ (15000 ${\rm M}_\odot$ for m15) and are precluded from de-refinement (merging into a regular gas element) until decelerating to 10\% of the launch velocity. Two particles are simultaneously spawned in opposite z-directions when the accumulated jet mass flux reaches twice the targeted spawned particle mass, ensuring exact conservation of linear momentum. Initially, the spawned particle is randomly positioned on a sphere with a radius of $r_0$, where $r_0$ is either 10 pc or half the distance between the black hole and the nearest gas particle, whichever is smaller. If the particle is initialized at coordinates $(r_0,\theta_0,\phi_0)$ and the jet opening angle for a specific model is $\theta{\rm op}$, the initial velocity direction of the jet is set at $2\theta_{\rm op}\theta_0/\pi$ for $\theta_0<\pi/2$ and at $\pi-2\theta_{\rm op}(\pi-\theta_0)/\pi$ for $\theta_0>\pi/2$. This ensures that the projected paths of any two particles do not intersect. A uniform value of $\theta_0=1^o$ is used for all simulations in this study.

The naming of each model starts from the halo model with different mass, the {\it primary} form of energy at our injection scale (`Kin',  `Th', and  `CR' for kinetic, thermal, and CR energy, respectively), and the corresponding energy flux in erg s$^{-1}$.`pr' at the end labels the model that is precessing. With initialized the jet magnetic fields in a toroidal form with a maximum field strength of $10^{-4}$ or $10^{-3}$ G, which all results in a sub-dominant magnetic energy flux.

\subsubsection{The jet models}
We concluded in \citetalias{2021MNRAS.507..175S} for a halo mass of $10^{14} M_\odot$ that the criteria for a successful jet model would generate a wide jet cocoon with sufficiently long cooling time. To fulfill these criteria, we found that the more favorable models invoke either a high-temperature thermal jet, a CR dominant jet, or a widely precessing kinetic jet. We, therefore focus on these three models and explore them in different halo masses. We describe each of the models and how we scale the energy input according to the halo mass below. 
A full list of simulations can be found in \tref{tab:run}. We emphasize that the parameters in the table reflect the jet parameters at our launch scale. The jet cocoon properties will continuously evolve as it interacts with the surrounding gas.
\begin{itemize}
   \item {\bf No jet:\,\,\,} These simulations are conducted without any jets but solely using the standard FIRE-2 stellar feedback, star formation, and cooling physics. They serve as a baseline testing the effect of any AGN jet.
   
    \item {\bf High-temperature thermal jet:\,\,\,} One method to have a jet cocoon with a sufficiently long cooling time is launching the jet at a high enough temperature so that the cooling time will be longer than the cocoon expansion time to the cooling radius. Given that the specific thermal energy (or temperature) is key for the cooling time, we keep a constant initial jet temperature of $T=3\times 10^9 K$ across the halo mass. The jet velocity $V_{\rm Jet}$ is set to be 3000 km s$^{-1}$, so the kinetic energy flux is subdominant. The dominant thermal component can also result in a faster lateral cocoon expansion that keeps the jet cocoon wide enough at the cooling radius to suppress a wide solid angle of cooling flows \citep{2021MNRAS.507..175S}. To a certain extent, this bears resemblance to the EAGLE AGN model \citep{2015MNRAS.446..521S}, particularly in terms of specific heat, albeit we have more collimated injection.

    \item {\bf CR dominant jet:\,\,\,} Another possibility to have a wide cocoon at the cooling radius with a sufficiently long cooling time is launching a CR-dominant jet. We keep the specific CR energy the same as the high-temperature thermal jet model, negligible thermal energy, and a subdominant jet velocity at 3000 km s$^{-1}$.
    \item {\bf Widely precessing kinetic jet:\,\,\,} Yet another way to have a sufficiently wide jet cocoon at the cooling radius is initially having a wide open angle or have jet precessing with a wide angle. In \citetalias{2021MNRAS.507..175S}, we showed that a jet with $\sim$ 100 Myr precessing period and a 45$^o$ angle can most efficiently and stably quench. We inherit that and have a jet velocity of 10000 $\rm{km\, s}^{-1}$, the same specific energy as the other two models.
\end{itemize}

\subsubsection{Scaling of AGN Jet as a function of halo mass}
In \citetalias{2021MNRAS.507..175S}, we found the above jet models with an energy flux of $10^{43}$ erg s$^{-1}$ can quench m14 systems. We therefore, use that as a reference point to anchor the jet energy fluxes for other halo masses. We test the following methods to scale the energy fluxes. Note that we use a constant flux over time.
\begin{itemize}
\item{\bf Free fall energy at $R_{\rm cool}$ (Sc-FF):\,\,\,} Given that the specific energy was concluded to be more relevant to the quenching criteria, we scale the energy flux with halo models by changing the mass fluxes preserving the specific energy of the jet. In \citetalias{2021MNRAS.507..175S}, we concluded the required jet energy flux is roughly comparable to the free-fall energy flux of gas at the cooling radius. We, therefore, use the jet energy flux of m14 as a reference point and scale it with the free-fall energy flux of each system ($\dot{E}_{\rm jet}\propto \dot{M}_{\rm cooling}v_{\rm ff}^2|_{r_{cool}}$). Here we define the cooling radius to be a radii beyond the cooling time is longer than 1.5 Gyr. The $\dot{M}{\rm cooling}$ was assessed based on the gas inflow of the no-jet run over a period of approximately 1-2 Gyr (0.2 Gyr for m15 due to its relatively long run time). Here, $v{\rm ff}$ represents the free fall velocity at the cooling radii.

\item{\bf Total cooling rate within $R_{\rm cool}$ (Sc-Cooling):\,\,\,} We also tried another way of scaling the jet energy flux (from the m14 runs), this time varying it in proportion to the total cooling rate within the cooling radius. The cooling rate is estimated by summing the radiative losses after running the no-jet run for 10 Myr. ($\dot{E}_{\rm jet}\propto \dot{E}_{\rm cool}|_{r<r_{cool}}$)
\end{itemize}
For m13l and m13, the second method results in a higher energy flux than the first. For m12 the above two methods result in very similar fluxes, so we included another set of runs with 10\% of the calculated m12 energy flux (Sc-0.1FF). We will further discuss the physical difference between the two scaling methods in later sections.

\setlength{\tabcolsep}{4pt}
\begin{table*}
\begin{center}
 \caption{Physics variations (run at highest resolution) in our halo-{\bf m14} survey}
 \label{tab:run}
\resizebox{17.7cm}{!}{%
\begin{tabular}{ccc|cc|ccc|ccc|cc|cc}
\hline
\hline
\multicolumn{3}{c|}{}&\multicolumn{2}{c|}{Results}&\multicolumn{6}{c|}{Input Jet Fluxes}&\multicolumn{4}{c}{Other Jet Parameters} \\
\hline
Model         &Scaling  &$\Delta t$& SFR &Summary  &$\dot{E}_{\rm Kin}$ &$\dot{E}_{\rm Th}$  &$\dot{E}_{\rm CR}$ &$\dot{M}$ &v &$\dot{P}$ &T & B &$\theta_{\rm p}$ & $T_{p}$ \\
               & &Gyr  &\multicolumn{2}{l|}{${\rm M}_\odot$ yr$^{-1}$}&\multicolumn{3}{c|}{erg s$^{-1}$} & ${\rm M}_\odot$ yr$^{-1}$ &km s$^{-1}$ &cgs &K &G & deg & {\rm Myr}\\
\hline

\hline
\hline 
\multicolumn{1}{c}{\bf \hlt{m12}}&&&\multicolumn{2}{c|}{}&\multicolumn{3}{c|}{}&\multicolumn{3}{c|}{}&\multicolumn{2}{c|}{}&\multicolumn{2}{c}{}\\
m12-NoJet             &N/A  &1.5 & 4.8 & strong CF &\multicolumn{3}{c|}{N/A} &\multicolumn{3}{c|}{N/A} &\multicolumn{2}{c|}{N/A} &\multicolumn{2}{c}{N/A} \\ 
{\bf m12-Th5e42}          &{\bf FF} &{\bf 1.5} & {\bf 0.031} & {\bf quenched} &{\bf 4.9e41} &\hlt{\bf 4.9e42}       &{\bf 0}      &{\bf 0.17} &{\bf 3e3}   &{\bf 3.3e33} &{\bf 3e9}  &{\bf 1e-4} &\multicolumn{2}{c}{\bf N/A} \\ 
{\bf m12-Th5e41}          &{\bf 0.1FF}&{\bf 1.5} & {\bf 0.67} & {\bf quenched } &{\bf 4.9e40} &\hlt{\bf 4.9e41}       &{\bf 0}      &{\bf 0.017} &{\bf 3e3}   &{\bf 3.3e32} &{\bf 3e9}  &{\bf 1e-4} &\multicolumn{2}{c}{\bf N/A} \\ 
{\bf m12-CR5e42}          &{\bf FF}&{\bf 1.5} &{\bf 0.016} & {\bf quenched} &{\bf 4.9e41} &{\bf 1.6e40}       &\hlt{\bf 4.9e42} &{\bf 0.17} &{\bf 3e3}   &{\bf 3.3e33} &{\bf 1e7}  &{\bf 1e-4} &\multicolumn{2}{c}{\bf N/A}  \\ 
{\bf m12-CR5e41}      &{\bf 0.1FF}    &{\bf 1.5} &{\bf 0.29} & {\bf quenched} &{\bf 4.9e40} &{\bf 1.6e39}       &\hlt{\bf 4.9e41} &{\bf 0.017} &{\bf 3e3}   &{\bf 3.3e32} &{\bf 1e7}  &{\bf 1e-4} &\multicolumn{2}{c}{\bf N/A}  \\ 
{\bf m12-Kin5e42-pr} &{\bf FF} &{\bf 1.5} &{\bf 0.11} & {\bf quenched} &\hlt{\bf 4.9e42} &{\bf 1.6e40}      &{\bf 0}      &{\bf 0.17} &{\bf 9.5e3} &{\bf 1.e34} &{\bf 1e7}  &{\bf 1e-4} &{\bf 45} &{\bf 100}\\ 
{ m12-Kin5e41-pr} &0.1FF&{ 1.5} &{ 1.38} & { strong CF } &\hlt{ 4.9e41} &{ 1.6e39}      &{ 0}      &{ 0.017} &{ 9.5e3} &{ 1.e33} &{ 1e7}  &{ 1e-4} &{ 45} &{ 100}\\ 
\hline
\multicolumn{1}{c}{\bf \hlt{m13 light}}&&&\multicolumn{2}{c|}{}&\multicolumn{3}{c|}{}&\multicolumn{3}{c|}{}&\multicolumn{2}{c|}{}&\multicolumn{2}{c}{}\\
m13l-NoJet              &N/A &1.5 & 3.0 & strong CF &\multicolumn{3}{c|}{N/A} &\multicolumn{3}{c|}{N/A} &\multicolumn{2}{c|}{N/A} &\multicolumn{2}{c}{N/A} \\ 
{\bf m13l-Th8e42}          &{\bf Cooling}&{\bf 1.5} & {\bf 0.032} & {\bf quenched} &{\bf 7.7e41} &\hlt{\bf 7.7e42}       &{\bf 0}      &{\bf 0.27} &{\bf 3e3}   &{\bf 5.2e33} &{\bf 3e9}  &{\bf 1e-4 (t)} &\multicolumn{2}{c}{\bf N/A} \\ 
{\bf m13l-Th1e42}          &{\bf FF}&{\bf 1.5} & {\bf 0.26} & {\bf quenched} &{\bf 1.3e41} &\hlt{\bf 1.3e42}       &{\bf 0}      &{\bf 0.044} &{\bf 3e3}   &{\bf 8.5e32} &{\bf 3e9}  &{\bf 1e-4 (t)} &\multicolumn{2}{c}{\bf N/A} \\ 
{m13l-CR8e42}          &{ Cooling}&{ 1.5} &0 & { overheated} &{ 7.7e41} &{ 2.5e40}       &\hlt{ 7.7e42} &{ 0.27} &{ 3e3}   &{ 5.2e33} &{ 1e7}  &{ 1e-4 (t)} &\multicolumn{2}{c}{ N/A}  \\ 
{\bf m13l-CR1e42}          &{\bf FF}&{\bf 1.5} &{\bf 0} & {\bf quenched} &{\bf 1.3e41} &{\bf 4.1e39}       &\hlt{\bf 1.3e42} &{\bf 0.044} &{\bf 3e3}   &{\bf 8.5e32} &{\bf 1e7}  &{\bf 1e-4 (t)} &\multicolumn{2}{c}{\bf N/A} \\ 
{m13l-Kin8e42-pr} &{Cooling}&{ 1.5} &{ 0} & { overheated} &\hlt{ 7.7e42} &{ 2.5e40}      &{ 0}      &{ 0.27} &{9.5e3} &{1.6e34} &{ 1e7}  &{ 1e-4 (t)} &{ 45} &{ 100}\\ 
{\bf m13l-Kin1e42-pr} &{\bf FF}&{\bf 1.5} &{\bf 0.12} & {\bf quenched} &\hlt{\bf 1.3e42} &{\bf 4.1e39}      &{\bf 0}      &{\bf 0.044} &{\bf 9.5e3} &{\bf 2.6e33} &{\bf 1e7}  &{\bf 1e-4 (t)} &{\bf 45} &{\bf 100}\\ 
\hline
\multicolumn{1}{c}{\bf \hlt{m13}}&&&\multicolumn{2}{c|}{}&\multicolumn{3}{c|}{}&\multicolumn{3}{c|}{}&\multicolumn{2}{c|}{}&\multicolumn{2}{c}{}\\
m13-NoJet              &N/A &1.5 & 7.0 & strong CF &\multicolumn{3}{c|}{N/A} &\multicolumn{3}{c|}{N/A} &\multicolumn{2}{c|}{N/A} &\multicolumn{2}{c}{N/A} \\ 
{\bf m13-Th7e42}        & {\bf Cooling}  &{\bf 1.5} & {\bf 0.060} & {\bf quenched} &{\bf 7.4e41} &\hlt{\bf 7.4e42}       &{\bf 0}      &{\bf 0.26} &{\bf 3e3}   &{\bf 5.0e33} &{\bf 3e9}  &{\bf 1e-4 (t)} &\multicolumn{2}{c}{\bf N/A} \\ 
{\bf m13-Th3e42}        &{\bf FF}  &{\bf 1.5} & {\bf 0.10} & {\bf quenched} &{\bf 3.3e41} &\hlt{\bf 3.3e42}       &{\bf 0}      &{\bf 0.12} &{\bf 3e3}   &{\bf 2.2e33} &{\bf 3e9}  &{\bf 1e-4 (t)} &\multicolumn{2}{c}{\bf N/A} \\ 
{ m13-CR7e42}        & {Cooling}  &{ 1.5} &0 & { overheated} &{ 7.4e41} &{ 2.4e40}       &\hlt{ 7.4e42} &{ 0.26} &{ 3e3}   &{ 5.0e33} &{ 1e7}  &{ 1e-4 (t)} &\multicolumn{2}{c}{N/A}  \\ 
{\bf m13-CR3e42}        &{\bf FF}  &{\bf 1.5} &{\bf 0} & {\bf quenched} &{\bf 3.3e41} &{\bf 1.1e40}       &\hlt{\bf 3.3e42} &{\bf 0.12} &{\bf 3e3}   &{\bf 2.2e33} &{\bf 1e7}  &{\bf 1e-4 (t)} &\multicolumn{2}{c}{\bf N/A}  \\ 
{ m13-Kin7e42-pr} & {Cooling} &{ 1.5} &{ 0} & { overheated} &\hlt{ 7.4e42} &{2.4e40}      &{ 0}      &{0.26} &{9.5e3} &{1.5e34} &{1e7}  &{1e-4 (t)} &{ 45} &{ 100}\\ 
{\bf m13-Kin3e42-pr} &{\bf FF}&{\bf 1.5} &{\bf 0.24} & {\bf quenched} &\hlt{\bf 3.3e42} &{\bf 1.1e40}      &{\bf 0}      &{\bf 0.12} &{\bf 9.5e3} &{\bf 6.8e33} &{\bf 1e7}  &{\bf 1e-4 (t)} &{\bf 45} &{\bf 100}\\ 
\hline
\multicolumn{1}{c}{\bf \hlt{m13 massive BH}}&&&\multicolumn{2}{c|}{}&\multicolumn{3}{c|}{}&\multicolumn{3}{c|}{}&\multicolumn{2}{c|}{}&\multicolumn{2}{c}{}\\
m13-mBH-NoJet             &N/A  &1.5 & 2.9 & strong CF &\multicolumn{3}{c|}{N/A} &\multicolumn{3}{c|}{N/A} &\multicolumn{2}{c|}{N/A} &\multicolumn{2}{c}{N/A} \\ 
{\bf m13-mBH-Th1e43}         &{\bf Cooling} &{\bf 1.5} & {\bf 0.067} & {\bf quenched} &{\bf 1.1e42} &\hlt{\bf 1.1e43}       &{\bf 0}      &{\bf 0.38} &{\bf 3e3}   &{\bf 7.3e33} &{\bf 3e9}  &{\bf 1e-4 (t)} &\multicolumn{2}{c}{\bf N/A} \\ 
{\bf m13-mBH-Th1e42}         & {\bf FF} &{\bf 1.5} & {\bf 0.2} & {\bf quenched} &{\bf 1.3e41} &\hlt{\bf 1.3e42}       &{\bf 0}      &{\bf 0.047} &{\bf 3e3}   &{\bf 9.0e32} &{\bf 3e9}  &{\bf 1e-4 (t)} &\multicolumn{2}{c}{\bf N/A} \\ 
{\bf m13-mBH-CR1e43}         & {\bf Cooling} &{\bf 1.5} &{\bf 0.} & {\bf quenched} &{\bf 1.1e42} &{\bf 3.5e40}       &\hlt{\bf 1.1e43} &{\bf 0.38} &{\bf 3e3}   &{\bf 7.3e33} &{\bf 1e7}  &{\bf 1e-4 (t)} &\multicolumn{2}{c}{\bf N/A}  \\ 
{m13-mBH-CR1e42}         & {FF} &{1.5} &2.5 & {strong CF} &{1.3e41} &{4.4e39}       &\hlt{1.3e42} &{0.047} &{3e3}   &{9.0e32} &{1e7}  &{1e-4 (t)} &\multicolumn{2}{c}{N/A}  \\ 
{\bf m13-mBH-Kin1e43-pr} & {\bf Cooling}&{\bf 1.5} &{\bf 0.007} & {\bf quenched} &\hlt{\bf 1.1e43} &{\bf 3.5e40}      &{\bf 0}      &{\bf 0.38} &{\bf 9.5e3} &{\bf 2.2e34} &{\bf 1e7}  &{\bf 1e-4 (t)} &{\bf 45} &{\bf 100}\\ 
{m13-mBH-Kin1e42-pr} & {FF}&{1.5} &{1.1} & {strong CF} &\hlt{1.3e42} &{4.4e39}      &{0}      &{ 0.047} &{9.5e3} &{2.8e33} &{ 1e7}  &{ 1e-4 (t)} &{ 45} &{ 100}\\ 
\hline
\multicolumn{1}{c}{\bf \hlt{m14}}&&&\multicolumn{2}{c|}{}&\multicolumn{3}{c|}{}&\multicolumn{3}{c|}{}&\multicolumn{2}{c|}{}&\multicolumn{2}{c}{}\\
m14-NoJet              &N/A &1.5 & 51 & strong CF &\multicolumn{3}{c|}{N/A} &\multicolumn{3}{c|}{N/A} &\multicolumn{2}{c|}{N/A} &\multicolumn{2}{c}{N/A} \\ 
{\bf m14-Th6e43}        &{\bf reference}  &{\bf 1.5} & {\bf 0.88} & {\bf quenched} &{\bf 5.8e42} &\hlt{\bf 5.8e43}       &{\bf 0}      &{\bf 2.0} &{\bf 3e3}   &{\bf 3.9e34} &{\bf 3e9}  &{\bf 1e-3 (t)} &\multicolumn{2}{c}{\bf N/A} \\ 
{\bf m14-CR6e43}        &{\bf reference}  &{\bf 1.5} &{\bf 0.16} & {\bf quenched} &{\bf 5.8e42} &{\bf 1.9e41}       &\hlt{\bf 5.8e43} &{\bf 2.0} &{\bf 3e3}   &{\bf 3.9e34} &{\bf 1e7}  &{\bf 1e-3 (t)} &\multicolumn{2}{c}{\bf N/A}  \\ 
{\bf m14-Kin6e43-pr} &{\bf reference} &{\bf 1.5} &{\bf 1.1} & {\bf quenched} &\hlt{\bf 5.8e43} &{\bf 1.9e41}      &{\bf 0}      &{\bf 2.0} &{\bf 9.5e3} &{\bf 1.2e35} &{\bf 1e7}  &{\bf 1e-3 (t)} &{\bf 45} &{\bf 100}\\ 
\hline
\multicolumn{1}{c}{\bf \hlt{m15}}&&&\multicolumn{2}{c|}{}&\multicolumn{3}{c|}{}&\multicolumn{3}{c|}{}&\multicolumn{2}{c|}{}&\multicolumn{2}{c}{}\\
m15-NoJet              &N/A &1.5 & 770 & strong CF &\multicolumn{3}{c|}{N/A} &\multicolumn{3}{c|}{N/A} &\multicolumn{2}{c|}{N/A} &\multicolumn{2}{c}{N/A} \\ 
{\bf m15-Th7e44}        & {\bf FF}  &{\bf 1.} & {\bf 85} & {\bf strong $\downarrow$} &{\bf 7.2e43} &\hlt{\bf 7.2e44}       &{\bf 0}      &{\bf 26} &{\bf 3e3}   &{\bf 4.9e35} &{\bf 3e9}  &{\bf 1e-4 (t)} &\multicolumn{2}{c}{\bf N/A} \\ 
{\bf m15-CR7e44}        & {\bf FF}  &{\bf 1.} &{\bf 59} & {\bf strong $\downarrow$} &{\bf 7.2e43} &{\bf 2.4e42}       &\hlt{\bf 7.2e44} &{\bf 26} &{\bf 3e3}   &{\bf 4.9e35} &{\bf 1e7}  &{\bf 1e-4 (t)} &\multicolumn{2}{c}{\bf N/A}  \\ 
\hline 
\hline
\end{tabular}
}
\end{center}
\begin{flushleft}
This is a partial list of simulations studied here: each halo was run with jet models and fluxes scaled from the more successful m14 run we concluded in \citetalias{2021MNRAS.507..175S}. Columns list: 
(1) Model name:  The naming of each model starts with the initial condition. The number following 'm' represents the logarithmic halo mass. Subsequently, it indicates the primary energy form, followed by the energy flux. 'pr' labels precession.
(2) Scaling: We scaled the energy flux by changing the mass flux while keeping the specific energy. The m14 runs are the reference point, so they are labeled as 'reference'. We scale either according to the free-fall energy within $R_{\rm cool}$ (`FF') or the total cooling rate within $R_{\rm cool}$ (`Cooling'). For m12, the 2 models yield similar energy fluxes so we did another with 0.1 that energy flux.
(3) $\Delta t$: Simulation duration. All simulations are run to $1.5\,$Gyr, unless either the halo is completely ``blown out'' or completely unaffected. 
(4) The SFR averaged over the last 250 Myr.
(5) Summary of the results. `strong CF', and `quenched' correspond respectively to an sSFR of $\gtrsim 10^{-11}$,  and $\lesssim 10^{-11} {\rm yr}^{-1}$. `Strong $\downarrow$' means at least one-dex suppression of SFR compared to the `NoJet' run. `Overheated' means the jet explosively destroys the cooling flow in $<500$ Myr, leaving a core with much lower density and high entropy and temperature (e.g. $\gg10^9$ K), in tension with observational constraints (detailed in \sref{s:gas}).
(6) $\dot{E}_{\rm Kin}$, $\dot{E}_{\rm Th}$, $\dot{E}_{\rm Mag}$, and $\dot{E}_{\rm CR}$ tabulate the total energy input of the corresponding form. The dominant energy form is highlighted in blue.
(7) $\dot{M}$, v, and $\dot{P}$ tabulate the mass flux, jet velocity, and momentum flux.
(8) T: The initial temperature of the jet.
(9) B: The maximum initial magnetic field strength of the jet; (t) and (p) means toroidal and poloidal respectively.
(10) $\theta_{p}$: The precession angle of the jet.
(11)  $T_{p}$: Precession period. 
\end{flushleft}
\end{table*}
\setlength{\tabcolsep}{6pt}


\section{Results} \label{S:results}
\subsection{Jet cocoon morphology}
\begin{figure*}\label{fig:morph}
    \centering
    \includegraphics[width=16cm]{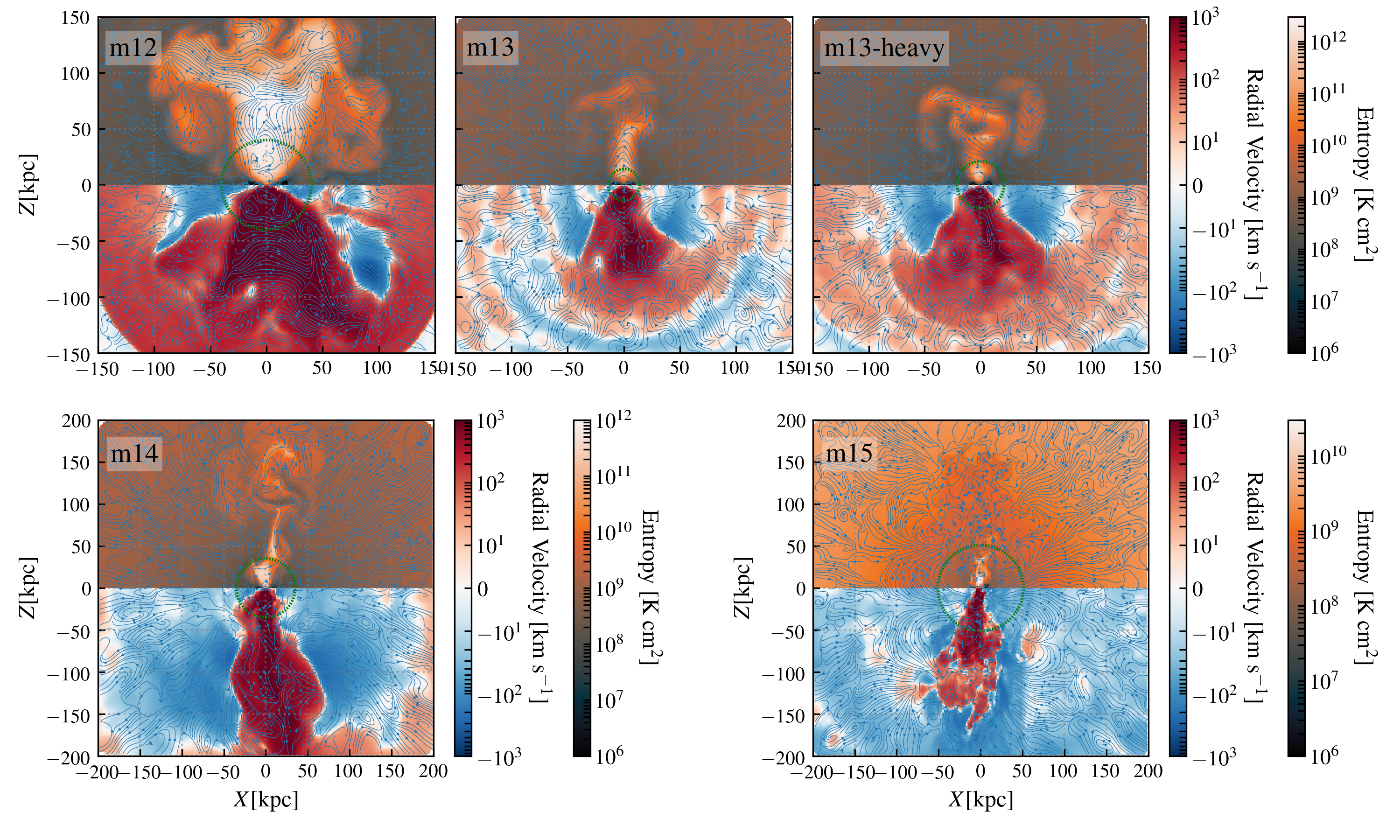}
    \caption{The entropy (upper) and radial velocity (lower) for a central slice of the cosmic ray jet run are displayed for all halo masses, with energy fluxes scaled to the free-fall energy fluxes at the cooling radii (``FF''). The magnetic field lines are overlaid on top. The plot represents the state at 500 Myr after the start of the run. As observed in the radial velocity panel, the jet cocoons exhibit quasi-isotropic behavior at the cooling radii (green circle) for each halo. The m15 runs depict a weaker bubble, consistent with the later growth of cooling flows. Magnetic fields are highly entangled.}
    \label{fig:res_acc}
\end{figure*}

In \fref{fig:morph}, the upper panel illustrates entropy, while the lower panel depicts radial velocity for a central slice of the cosmic ray jet run across various halo masses with an energy flux scaled relative to the free-fall energy fluxes at the cooling radii. Overlaid on the plot are magnetic field lines. 

Notably, in the radial velocity panel, the jet cocoons display quasi-isotropic characteristics at the cooling radii, marked by green circles, within each halo. The m15 runs exhibit a comparably weaker bubble, aligning with the subsequent growth of cooling flows, which will be discussed later. The morphology shows a wider outflowing region than the high-entropy region in most cases, as the cosmic ray is more extended than the thermal energy.
Remarkably, the magnetic fields are intricately entangled.

\subsection{Star formation rate}
\begin{figure}\label{fig:sfr}
    \centering
    \includegraphics[width=8cm]{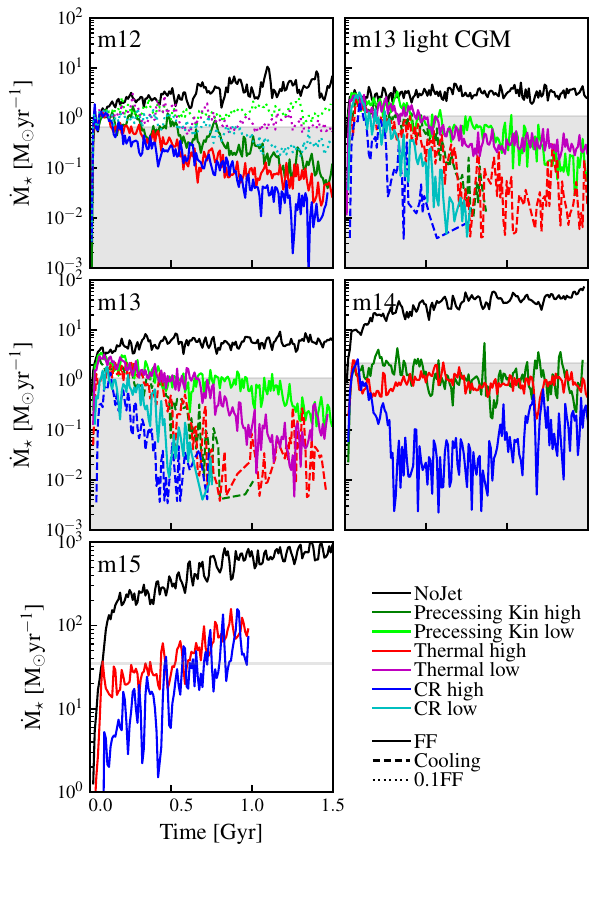}
    \caption{The star formation rate as a function of time. `Higher' and `Lower' label the models with higher and lower fluxes. The runs with jet energy flux scaled with the free-fall energy at the cooling radius ($R_{\rm cool}$) or total cooling rate within $R_{\rm cool}$ are the solid and dashed lines, respectively. The dotted line for m12 is the run with 0.1 the energy flux yielded from scaling with the free-fall energy at $R_{\rm cool}$. The shaded region for m12, m13l, m13, and m14 cover the region with a specific SFR (sSFR) of $10^{-11}$, which we defined as quenched. For m15, the gray line indicates an SFR of $30 M_{\odot}$, which is the upper bound NGC 1275, the BCG of the Perseus Cluster.
    The energy required to quench a halo roughly scales with the free-fall energy at the cooling radius. In all cases, with the same energy fluxes, CR jet quenches more efficiently. }
    \label{fig:res_acc}
\end{figure}

\fref{fig:sfr} shows the star formation rate of all the simulations we ran.  The runs with jet energy flux scaled with the free-fall energy at the cooling radius ($R_{\rm cool}$) (method 1 above) or the total cooling rate within $R_{\rm cool}$ (method 2 above) are shown with solid and dashed lines, respectively. The dotted line for m12 is the run with just 10\% of the energy flux as computed by scaling with the free-fall energy at $R_{\rm cool}$. 
The shaded region for m12, m13l, m13, and m14 cover the region with a specific SFR (sSFR) of $10^{-11}$, which we defined as quenched. For m15, the gray line indicates a SFR of $30 {\rm M_{\odot}  yr}^{-1}$, which is the upper bound for NGC 1275 \citep[e.g.,][]{1996AJ....111..130D,2010MNRAS.405..115C}, the BCG of the Perseus Cluster.
The energy required to quench a halo roughly scales with the free-fall energy at the cooling radius.  However, the resulting SFR in m15 is still a factor of 2-3 higher than the brightest central galaxy seen in Perseus. 
This discrepancy largely stems from the estimation of the mass cooling flow ($\dot{M}_{\rm cool}$) for m15, which was derived from the initial 200 Myr of the m15 no-jet run due to its long runtime. Upon extending the m15 no-jet run to 1 Gyr, the average cooling flow is observed to be three times higher than during the initial 200 Myr. If this value were adopted for scaling purposes and the jet energy fluxes were increased by a factor of 3, the star formation rate of m15 runs with jet would likely align with the observed upper bounds.

In all cases, with the same energy fluxes, CR jets quench more efficiently than the corresponding runs with thermal or kinetic jets. CR jets can quench the m12 halo, even with an energy flux a factor of 10 lower than obtained by scaling from m14 (according to the free-fall energy at the cooling radius, method 1 above).  This is roughly consistent with what we saw in \citetalias{2021MNRAS.507..175S} for the m14 cases where CR jets can quench with $\sim 10^{42} {\rm erg s}^{-1}$, a factor of 10 lower than the values we use here. The reason for the more efficient quenching is mostly the extra CR pressure support and the change in thermal instability. We discuss those further and possible uncertainties from cosmic ray transport modeling in \sref{s:CR_vs_Th} and \sref{s:caveat}.

\subsection{Cooling flow properties}
\begin{figure}\label{fig:hot}
    \centering
    \includegraphics[width=8cm]{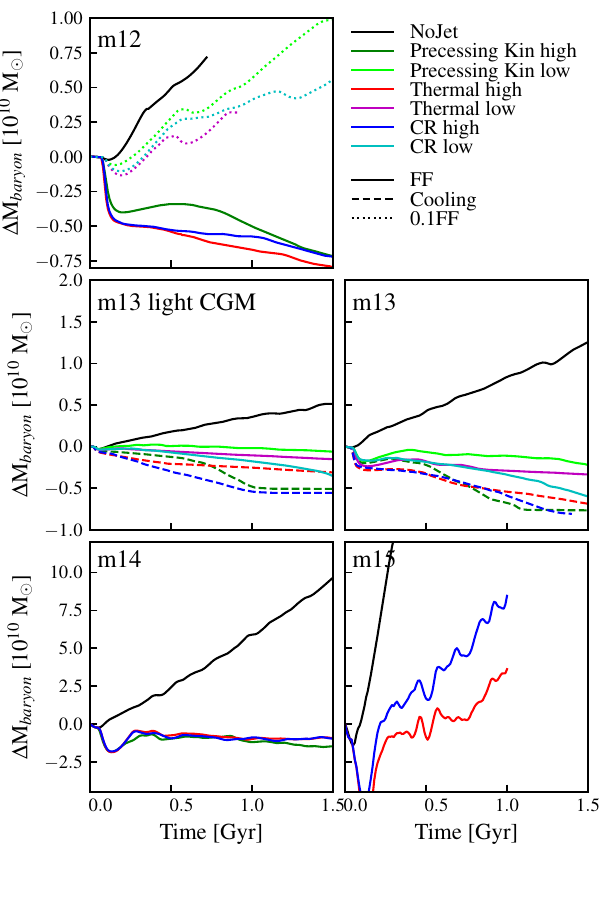}
    \caption{The change in baryonic mass within the cooling radius of each run. Runs with jet energy flux scaled according to the free-fall energy within the cooling radius have a roughly constant core baryonic mass throughout the simulation time, indicating roughly balanced cooling flows. But the core baryonic mass in the m15 run increases after 0.7 Gyr, consistent with the later rise of stellar mass. Runs with fluxes higher/lower than that result in a dropping/growing core baryonic mass.}
    \label{fig:res_acc}
\end{figure}

\fref{fig:hot} shows the baryonic mass within the cooling radius of each run. Runs with jet energy flux scaled according to the free-fall energy within the cooling radius (scaling method 1) have a roughly constant core baryonic mass throughout the simulation time, indicating roughly balanced cooling flows. This includes the higher energy flux runs for m12, the lower energy flux runs for m13l and m13, as well as m15. In m15, the core baryonic mass starts to rise after 0.7Gyr, consistent with the later rise of stellar mass. 

Run with energy fluxes higher (lower) than the scaled value results in a decreasing (increasing) core baryonic mass. Overall, with a similar energy flux, CR jet runs again suppress the cooling flow most efficiently.

\subsection{X-ray luminosity}
\fref{fig:xray} also shows the X-ray luminosity of each run over the last 100 Myr at the 0.5-7 keV band. We see little evolution of the X-ray luminosity for most of the runs except for the m13l and m12 runs with higher jet energy fluxes. All of the other runs have luminosities that differ from the initial condition by at most a factor of 2 and are within the observational range. This is true even without an AGN jet. 

The m12 and m13l runs with higher jet energy fluxes have slightly lower X-ray luminosities than the observed range. The stronger jet heats the gas, decreasing the gas density, which in turm lowers the X-ray luminosities. In particular, CR and thermal jets are more efficient in suppressing the X-ray fluxes.

\subsection{Gas profiles}\label{s:gas}

\begin{figure}\label{fig:etd_m12}
    \centering
    \includegraphics[width=8cm]{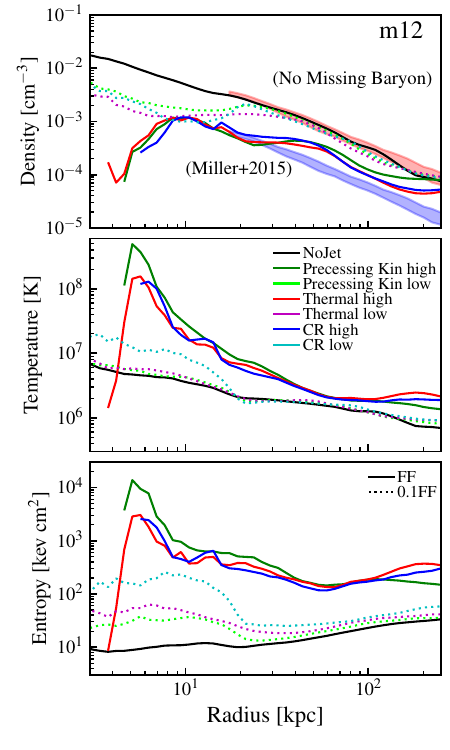}
    \caption{ The average gas density ({\em top row}), temperature ({\em second row}), and entropy profiles ({\em bottom row}) of the m12 runs are plotted against radius, representing an averaging over the most recent $\sim 100\,$Myr for the high-temperature gas ($>10^5$ K). We keep a separate plot for m12 as X-ray-luminosity-weighted profiles make less sense when the viral temperature is much less than the x-ray band. The shaded regions in the first row indicate observational density profiles for the Milky Way as outlined in \citet{2015ApJ...800...14M} (blue), accompanied by the same line scaled under the assumption of no missing baryons (red), constituting our initial condition. The higher-flux jets result in an overall lower density, higher temperature, and higher entropy, thereby progressing towards alignment with the observed outcomes for the Milky Way. However, they do not significantly alter the slops of the profiles. Different jet models with the same fluxes result in very similar gas profiles. The heated cores are all within 10 kpc.}
    \label{fig:res_acc}
\end{figure}

\begin{figure*}\label{fig:etd_all}
    \centering
    \includegraphics[width=16cm]{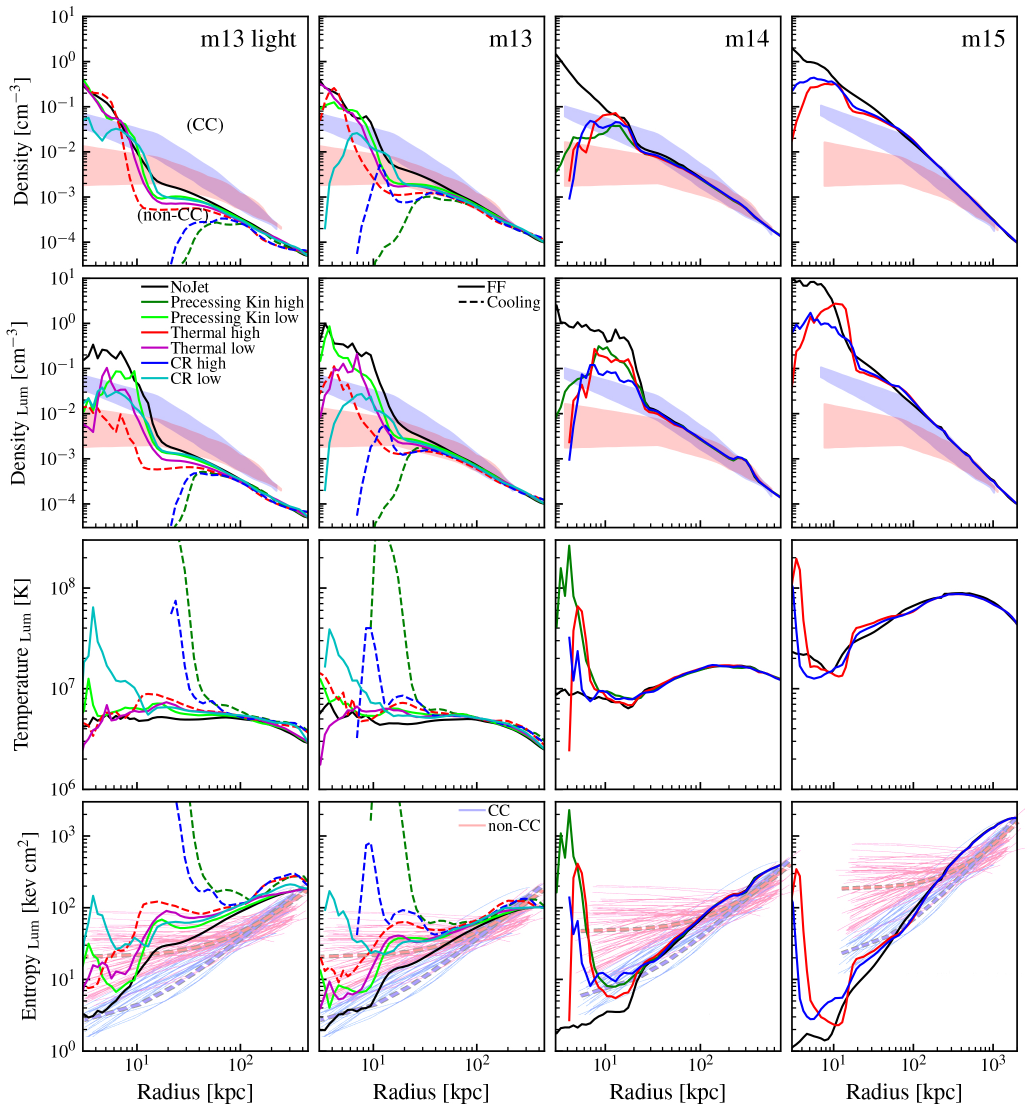}
    \caption{Mean gas  density ({\em top row}), X-ray cooling luminosity-weighted density ({\em second row}),  luminosity-weighted temperature ({\em third row}), and luminosity-weighted entropy ({\em bottom row}) versus radius averaged over the last $\sim 50\,$Myr in all the m13l, m13, m14, and m15 runs. 
    The shaded regions in the first and second row and the light curves in the bottom row indicate the observational density and entropy profiles (scaled) for cool-core (blue) and non-cool-core (red) clusters \citep{2013ApJ...774...23M} (scaled to account for the halo mass differences). Thermal and kinetic jet with higher energy fluxes results in a very large over-heated core in m13l and m13 runs, extending to 100 kpc scales. All the rest of the runs have gas profiles broadly consistent with the observational range and has heated core only limited within 10 kpc. }
    \label{fig:res_acc}
\end{figure*}

\fref{fig:etd_m12} and \fref{fig:etd_all} show the density, temperature, and entropy profiles of all the runs. Given the m12 viral temperature is an order of magnitude lower than the X-ray band, we use a mass-weighted, rather than X-ray-luminosity-weighted, profiles for gas hotter than $10^5$ K. For the halo mass $\gtrsim10^{13}{\rm M_\odot}$, we use an X-ray luminosity weighted value in the 0.5-7 keV band.  The shaded regions in \fref{fig:etd_m12} indicate  observational density profiles for the Milky Way as outlined in \cite{2015ApJ...800...14M} (blue), accompanied by the same line scaled under the assumption of no missing baryons (red), constituting our initial condition. The shaded regions in \fref{fig:etd_all} in the density and the entropy panels indicate the observational profiles (scaled) for cool-core (blue) and non-cool-core (red) clusters \citep{2013ApJ...774...23M} (scaled to account for the halo mass differences). Note that the shaded region is scaled from a halo mass of $\sim10^{14}{\rm M_\odot}$, so it will be most reasonably compared with m14 and m15.

In the m12 cases, when scaling the energy flux according to the free-fall energy at the cooling radius (the higher fluxes run), the resulting jet suppresses the gas density and heats the gas at all radii, resulting in a density profile  progressing towards alignment with the observed outcomes for the Milky Way.  The overall slope of the profiles remains similar within 100 kpc. With one dex lower fluxes, the resulting profiles roughly follow those of the `NoJet' runs with only a slightly raised entropy within 10 kpc. Different jet models with the same energy fluxes also result in very similar gas properties in m12, as the effective jet cocoon width and temperature are similar. 

For the heavier halos, jets with energy fluxes scaled with the free-fall energy at the cooling radius (the lower fluxes runs) have  much smaller impacts on the gas properties. The heated core region stays within 10 kpc in all cases. The more massive the halo is, the more the resulting profiles resemble the cool-core clusters. The less massive the halo is, the flatter the entropy profiles get.

For the m13 and m13l cases, if the energy fluxes are scaled according to the total cooling rate within the cooling radius (the 2-6 times higher fluxes runs), the resulting jets can much more significantly suppress the core density and heat up the gas. Among the 3 different models, the thermal jet is the only one that will result in more reasonable gas profiles. Both the CR and kinetic jets with higher energy fluxes result in very explosive feedback with heated cores and negative temperature gradients extending to $100$ kpc, resulting a larger tension with the observations.

\subsection{Turbulence induced by precessing jet}\label{s:turbulence}
\begin{figure*}\label{fig:turb}
    \centering
    \includegraphics[width=16cm]{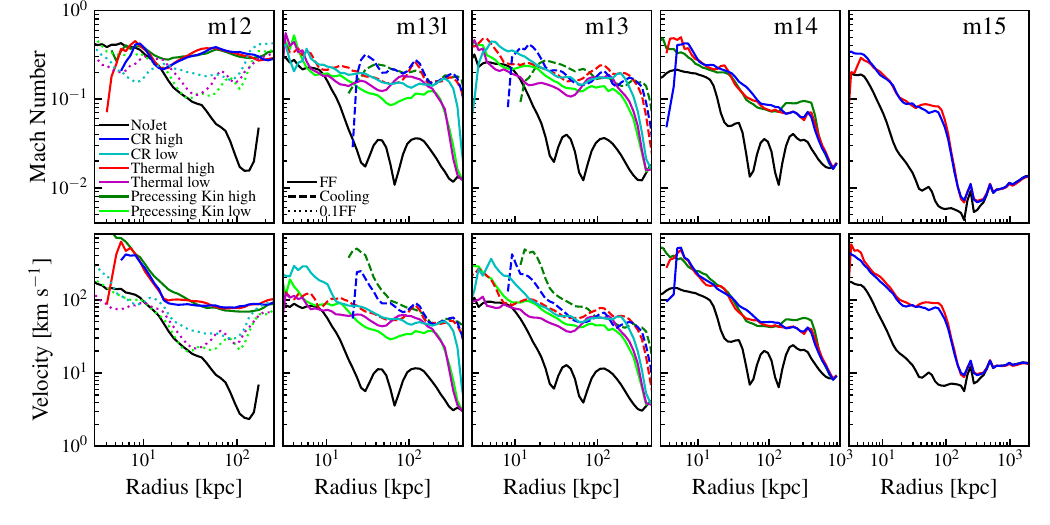}
    \caption{{\em Top row:} 1D rms Mach number in gas with $T>10^{6}\,$K (m12, m13l, m13) or $T>10^{7}\,$K (m14, m15) (averaged over the last 50\,Myr of the runs) as a function of radius for the runs. {\em Bottom row:} 1D rms velocity dispersion for the same gas. (We exclude the radial motion to obtain a rough upper bound on the turbulent velocity.)  Higher energy flux jets further boost the turbulent velocity at most radii.
    }
    \label{fig:res_acc}
\end{figure*}

All jet models increase the Mach number and turbulent velocity of the hot phase. In \fref{fig:turb}, we plot the turbulent velocity as a function of the radius of each run. For m12, m13l, and m13 runs, we define the hot gas to be $T>10^6$ K, while for more massive halos, we use a higher minimum temperature for the hot gas, $T>10^7$ K. The plotted turbulent velocity should be regarded as an upper bound, as we have only subtracted the velocity contributions of shell-averaged radial inflows and annuli-averaged rotations from the calculations. Scaling the energy flux according to the free-fall energy at the cooling radius results in the highest turbulent velocity for the m12 case, reaching 400-800 km s$^{-1}$. The maximum velocity reaches $\sim$400-500 km s$^{-1}$ for all of the other more massive cases. Of the halos we surveyed, the best observational constraint would be the m15 case, in which we obtain a maximum 1D Mach number of approximately 0.3, slightly higher than the observed upper bound of the Perseus cluster by a factor of 1.5-2 \citep{2016Natur.535..117H,2018PASJ...70....9H}. However, all of the highest velocities happen within 10 kpc.

The higher energy flux runs in m13l and m13 (Sc-Cooling) result in an even higher turbulent velocity at a large radius, reaching $\sim$400-500 km s$^{-1}$ at 20-30 kpc. The lowest energy flux runs in the m12 case (Sc-0.1FF) result in a lower turbulent velocity within 100 kpc, beyond which the velocity becomes similar to the other runs with higher flux jets. With a similar energy flux, the different jet models do not result in velocities significantly different from each other. 

\section{The quenching criteria as a function of halo mass}\label{s:criteria_mass}

\subsection{Revisiting the quenching criteria for a jet model}\label{s:criteria}
In \citetalias{2021MNRAS.507..175S}, we concluded that there were three major criteria that had to be met for a jet model to quench a $10^{14} {\rm M_\odot}$ halo. 
We review these criteria here for different halo masses.
\subsubsection{Sufficient energy flux}
We concluded that the jet energy flux should roughly scale with the free-fall energy at the cooling radius, consistent with what we stated in \citetalias{2021MNRAS.507..175S}. 
\begin{align}
\label{eq:e_min}
\dot{E}_{\rm min}&\sim\dot{M}_{\rm cool}v_{\rm ff}[R_{\rm cool}]^2\notag\\
&\sim10^{43}\, {\rm erg\,s}^{-1} \left(\frac{\dot{M}_{\rm cool}}{100\,{\rm {\rm M}_\odot\, yr}^{-1}}\right) \left(\frac{v_{\rm ff}[R_{\rm cool}]}{300\,{\rm km\, s}^{-1}}\right)^2.
\end{align}
We discuss the difference between scaling methods (Sc-FF and Sc-Cooling) in \sref{s:scaling}.

\subsubsection{Long cooling time in jet cocoon}\label{s:cocoon}
The second criterion is that the jet cocoon should have a long enough cooling time so that most of the energy does not radiate away before reaching the cooling radius. If the majority of the energy is in the form of cosmic rays, the energy does not radiate very efficiently. If the majority of the energy is in thermal form, the temperature has to be high enough to ensure a long enough cooling time to prevent rapid radiative losses.
Quantitatively, this means $t_{\rm cool}\sim kT/\bar{n}\Lambda(T)$ is much larger than $ t_{\rm exp}\sim R_{\rm cool}/v_{\rm exp}\propto\rho_R^{1/3}\dot{E}_{\rm tot,\,J}^{-1/3}R_{\rm cool}^{5/3}$, where $\Lambda$ is the cooling function.

Therefore, we have
\begin{align}\label{eq:temp}
T& > \frac{\bar{n}\Lambda(T)\rho_R^{4/3}R_{\rm cool}^{5/3}\dot{E}_{\rm tot,\,J}^{-1/3}}{k}\notag\\
&\sim 10^8 {\rm K} \left(\frac{\bar{n}}{0.01 {\rm cm}^{-3}}\right)^{7/3}\left(\frac{R_{\rm cool}}{30 {\rm kpc}}\right)^{5/3}\left(\frac{\dot{E}_{\rm tot,\,J}}{10^{45} {\rm erg\,s}^{-1}}\right)^{-1/3}.
\end{align}

The cooling radii for m12, m13l, m13, m14, and m15 are roughly 40, 14, 21, 35, and 51 kpc. We defined the cooling radius as the point where half of the gas has a cooling time longer than 1.5 Gyr, the maximum run time. We note that with a halo mass above $10^{13}{\rm M_\odot}$, the cooling radius is a monotonically increasing function of halo mass. For m12, the cooling radius is larger due to the lower viral temperature, which results in a shorter cooling time.

The density at the cooling radius of each case would roughly be  $10^{-3} {\rm cm}^{-3}$ for m12, m13l, and m13 cases ( \fref{fig:etd_m12} and \fref{fig:etd_all}). The density at the cooling radius is roughly $10^{-2}$ and $4\times10^{-2} {\rm cm}^{-3}$ for m14 and m15 cases. Putting in the densities into the right hand side of \Eqref{eq:temp}, the specific energy of our jet models corresponds to a temperature at least comparable to that is described in \Eqref{eq:temp}.

\subsubsection{Quasi-isotropic jet cocoon at $R_{\rm cool}$}\label{s:cocoon}
The jet cocoon also needs to be wide enough at the cooling radius, reaching semi-isotropic, so an order of unity of the solid angle can be covered.
With a precessing jet, such criteria are automatically fulfilled, as we started with a wide 45-degree precession. For thermal or cosmic ray jets, the fraction of kinetic energy in the jet should follow \citetalias{2021MNRAS.507..175S} as
\begin{align}
\label{eq:fkin_nonkin}
f_{\rm kin}\lesssim& 0.4 \left(\frac{v_{\rm J}}{3000\, {\rm km\,s}^{-1}}\right)^{1/2}\left(\frac{\dot{M}_{\rm J}}{2\,{\rm M}_\odot {\rm yr}^{-1} }\right)^{-1/2}\notag\\
          &\times \left(\frac{v_{\rm ff}}{300\, {\rm km\,s}^{-1} }\right)^{-1/2}
           \left(\frac{\dot{M}_{\rm cool}}{100\,{\rm M}_\odot {\rm yr}^{-1} }\right)^{1/2}.
\end{align}
When we scale the energy flux by changing the mass flux according to the free fall energy for different halo masses, we get
\begin{align}
\label{eq:fkin_nonkin}
f_{\rm kin}\lesssim& 0.4 \left(\frac{v_{\rm J}}{3000\, {\rm km\,s}^{-1}}\right)^{1/2} \left(\frac{v_{\rm ff}}{300\, {\rm km\,s}^{-1} }\right)^{-3/2}.
\end{align}
Both the thermal and CR jets have a kinetic fraction lower than that.

\subsection{Scaling for the jet energy flux}\label{s:scaling}
In this work, we tried two different scaling methods of jet energy in different halo masses. We scaled the energy flux of the successful jet model in m14 to other halo masses either according to the the free-fall energy flux at $R_{\rm cool}$  or total cooling rate within $R_{\rm cool}$. These two quantities are physically distinct as the following discussion makes clear.

The Mach number of the cooling flow scales with the free-fall time ($t_{\rm ff}$) and the cooling time ($t_{\rm cool}$) as \citep[see, e.g.,][]{2019MNRAS.488.2549S}
\begin{align}
&\mathcal{M}\equiv \frac{v_{flow}}{v_{sound}} \sim \frac{t_{ff}}{t_{cool}}.
\end{align}
We also know that the cooling time, free fall time, sound speed, and circular velocity roughly scale as
\begin{align}
&t_{cool}\sim \frac{N k T}{\Lambda n^2 V}\sim \frac{T}{\Lambda n},\,\,\,\,\,
t_{ff}\sim \frac{r}{v_{ff}}\notag\\
&v_{sound}\sim v_{ff}\sim v_{cir}\sim\sqrt{T}.
\end{align}

So the cooling flow velocity roughly follows
\begin{align}
&v_{flow}\sim\frac{r\Lambda n}{T^{0.5} v_{cir}}\sim r \Lambda n T^{-1}. 
\end{align}

The total mass and energy fluxes are, therefore,
\begin{align}
&\dot{M}_{flow}\sim 4\pi r^2 n v_{flow}\sim r^3\Lambda n^2 T^{-1} \notag\\
&\dot{E}_{flow}\sim \dot{M}_{flow} v_{ff}^2\sim r^3\Lambda n^2 \sim V \Lambda n^2,
\end{align}
where V is the volume within $r_{cool}$, and all the quantities above are evaluated at the $R_{\rm cool}$.

Whereas,
\begin{align}
\dot{E}_{cool} \sim \int\limits_{r<r_{cool}} dV \Lambda n^2.
\end{align}

The net difference between the two will depend on how the cooling rate within the cooling radius scales differently from the cooling rate at the cooling radius. In the m12 case, the two scaling relations from m14 give similar results. For m13 and especially in m13l, the two scalings can yield a factor of 2-6 difference. The reason is that m13 (and especially m13l) has a comparatively weaker cooling flow than the other halo masses. Its viral temperature is at the minimum of the cooling curve, and the density is not as high as the more massive halos. Despite the lower cooling rate at the cooling radius, at the very core region of the galaxy ($<5$ kpc), the cooling from the ISM is dominant. Given that one of the scalings is dominated by ISM cooling while the other scaling is not, the above two expressions give very different results.

Summarizing from the differences among m13, m13l and, m13-mBH, there are at least two things that will add to the difference between the two scalings. The first factor is the CGM mass. A lighter CGM will cool even less efficiently than the ISM gas. Since the core cooling is always dominated by the ISM, including ISM cooling or not in the expression makes a larger difference in the lighter CGM case.

The difference can be even larger if we also increase the black hole mass at the galaxy center. The more massive black hole makes the ISM gas even denser at small radii and its cooling rate even higher. However, this has no impact on the cooling rate at the cooling radius. 
We will discuss this further in \sref{s:BH_king} for the m13l-mBH runs, where the two different scalings give more than a dex difference in the estimated jet energy fluxes.   

We note that scaling the energy flux according to the free fall energy at the cooling radius, which does not account for the ISM cooling, gives a better result for stably suppressing the cooling flow. The reason is that the AGN jet is working against the bulk cooling flow, which is from the hot halo and has a longer-term variability. The ISM, on the other hand, has larger short-term variability, especially when part of the gas is heated up by the AGN. Once the AGN jet attains an energy flux sufficient to suppress the cooling flows, the supply of ISM is curtailed, causing ISM coolings to decline as the ISM is gradually consumed. The instantaneous strong ISM cooling is thus not something that the AGN jet needs to counterbalance over an extended timescale. Additionally, ISM cooling is partly compensated for by supernovae operating on a shorter timescale. Having a jet flux anchoring to the CGM cooling on a long time scale, therefore, yields a more stable result.

\subsection{Thermal vs CR jet}\label{s:CR_vs_Th}
In all the surveyed cases with the same energy flux, CR jets always suppress the star formation more efficiently. The major reason is again the extra CR pressure gradient, as we first described in \citetalias{2021MNRAS.507..175S}.This is also consistent with what we observed in \cite{2023MNRAS.520.5394W}.

\begin{figure*}\label{fig:pres}
    \centering
    \includegraphics[width=16cm]{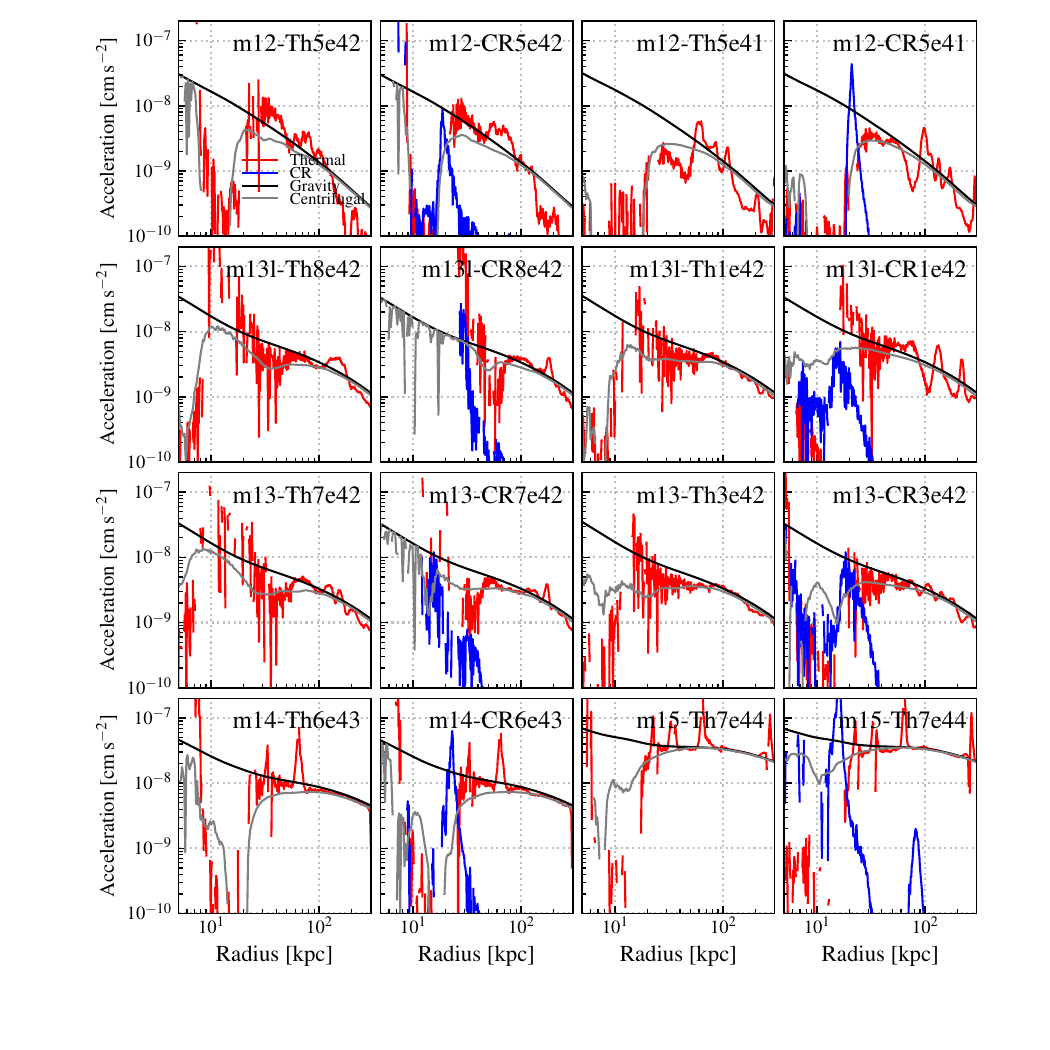}
    \caption{Comparison of gravitational, rotational, thermal pressure, and CR pressure gradient acceleration. The centrifugal acceleration is defined as $GM_{\rm enc}/r^2-v_{\rm rot}^2/r$. In the core region, where cooling is rapid, the thermal pressure gradient is not outward, and support is lost. In our CR jet runs, the CR pressure gradient predominantly balances gravity in the core region.}
    \label{fig:res_acc}
\end{figure*}

\fref{fig:pres} shows the centrifugal acceleration due to gravity and rotation and the acceleration due to the pressure gradient. Again at a few 10 kpc scales, the thermal radiative cooling is efficient, and the thermal pressure gradient is not always pointing outward. The CR pressure gradient, on the other hand, can reach a comparable value to the centrifugal acceleration at the same radius.

\subsection{Kinematic effect of BH mass}\label{s:BH_king}
In all the runs studied here, we explored jet models with fixed fluxes, which are not tied to black hole accretion or the exact black hole mass. However, the black hole mass still affects the surrounding gas, the star formation, and the propagation of the jet. To explore this effect, here we run a set of m13-mBH runs with a halo mass the same as m13 but a black hole mass of $5\times 10^9{\rm M_\odot}$, comparable to the black hole mass in the M87 halo. 

\begin{figure}\label{fig:bh_sf}
    \centering
    \includegraphics[width=8cm]{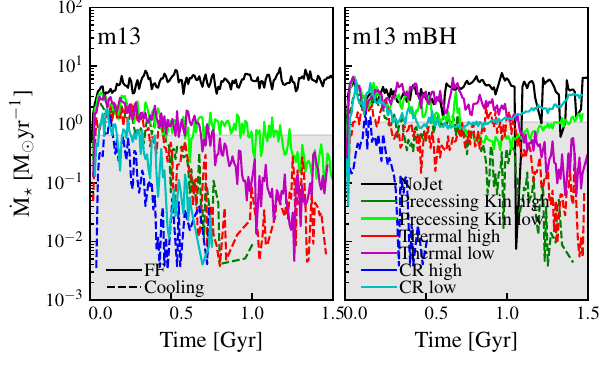}
    \caption{The star formation rate as a function of time for the m13 and m13-mBH runs. The more massive black hole in the m13-mBH run results in a more episodic star formation rate (in the run without jets) due to the concentrated star formation around the black hole. The more concentrated gas around the massive black hole also makes low-flux CR jets less efficient in quenching than the low-flux thermal jet.}
    \label{fig:res_acc}
\end{figure}

The resulting star formation rate is shown in \fref{fig:bh_sf}. The lower flux precessing kinetic jet and the CR jet runs which have energy fluxes scaled from m14 according to the free-fall energy at the cooling radius, both fail to quench. Initially, we see a similar suppression of the star formation rate, but later the star formation rate goes up again to several $M_\odot$ per year. A thermal jet with lower energy flux, on the other hand, can still effectively quench the galaxy. The primary reason is the kinematic effect of the black hole mass.

\begin{figure}\label{fig:star}
    \centering
    \includegraphics[width=8cm]{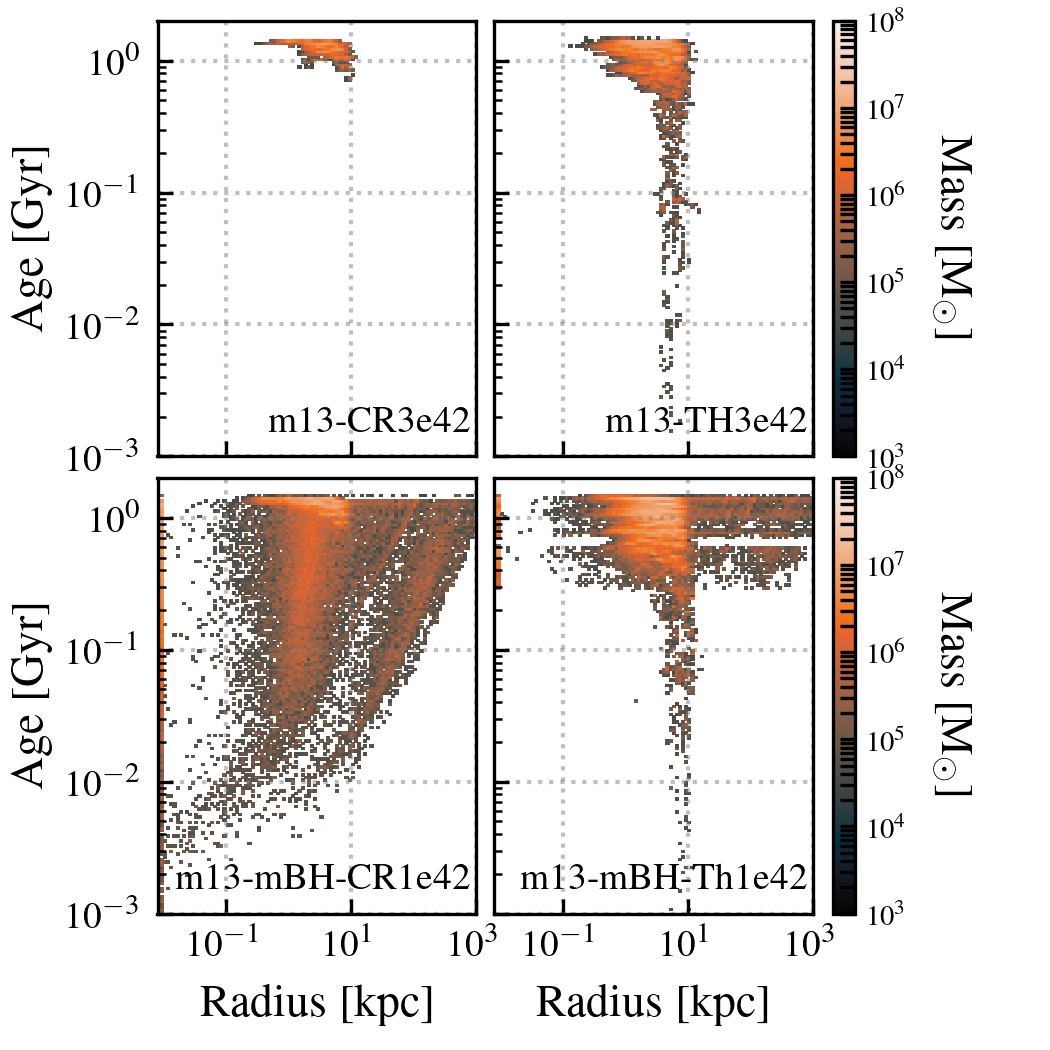}
    \caption{The age vs. radius of all the stars formed in the simulations for the low-flux thermal and CR jets in the m13 and m13-mBH cases. The inclusion of a massive black hole results in more dense gas and more concentrated star formation around the black hole vicinity. Cosmic ray dominated jets are less efficient in repelling the gas and regulating the star formation at the $<$100 pc scale.  }
    \label{fig:res_acc}
\end{figure}

The more massive black hole in the center has at least three kinematic effects. First of all, it results in much denser gas in the black hole vicinity, which can more efficiently form stars.
As we see in \fref{fig:star},  where the stellar age is plotted with respect to the radial position, the run with the more massive black hole and CR jet has most of the stars formed on $<100$ pc scales, especially at late times. Thermal jets can more effectively repel the gas at within 1 kpc, while CR jets cannot. As shown in \fref{fig:pres}, the extra CR pressure gradient mostly exists on the $\sim10$ kpc scale. At the sub-parsec scale, where the gas density is high, CRs also dissipate energy efficiently through hadronic and Coulomb losses. Due to the relatively high magnetic field strength in dense gas, the streaming of cosmic rays is also more efficient.

Another effect is the enhanced clustered star formation in the massive black hole run, which also results in a more episodic star formation due to the stellar cluster feedback. This can be very clearly seen in the SFR of the `NoJet' run in the massive black hole case.

Finally, the denser gas around the more massive black hole also alters the cooling rate in the ISM, enhancing it compared to the cooling rate at the cooling radius. This further enlarges the difference between the energy fluxes obtained from the two different scalings described in \sref{s:scaling}.

\section{Discussion}\label{s:discussion}
\subsection{Limitation of our models}
\label{s:caveat}

It is essential to underscore that our study deliberately examines jet models featuring a constant flux within a specific initial cluster configuration. Our model lacks dynamically-variable black hole accretion, making it non-``self-regulating.'' In scaling the energy flux based on the halo mass, we considered either the free fall energy at the cooling radius or the total cooling rate within the cooling radius, as computed from the `NoJet' run. We did not incorporate the evolution of these quantities after activating the jets.

The primary objective of this research is to identify the criteria (such as fluxes, energy forms, etc.) for a jet to induce quenching in various halo masses without overtly violating observational constraints. Observationally, in the m12 case, resembling an L* galaxy, there is  active star-forming populations. However, our focus in this paper is specifically on the quenching of star formation.

Another constraint of this study is the absence of a cosmological context in our simulations. Specifically, our halo lacks satellites/substructures, which could influence large-scale turbulence and other properties of the CGM/ICM gas. Additionally, the observed limited variation in X-ray luminosity across our runs results from initiating our isolated galaxy simulation with conditions resembling a cool-core cluster and running it for a duration of less than $1.5$ Gyr. In cosmological simulations, X-ray luminosity and other thermodynamic properties can be even more sensitive to the models of AGN feedback \citep[e.g.,][]{2010MNRAS.406..822M,2014MNRAS.441.1270L,2014MNRAS.438..195P,2015MNRAS.449.4105C,2016MNRAS.456.4266L,2018MNRAS.479.5385H,2019MNRAS.486.2827D}.

We assume a constant diffusivity for cosmic ray transport. The models for cosmic ray transport can introduce uncertainties in the resulting cosmic ray pressure profile. Simulations with variable diffusion, following self-confinement or extrinsic turbulence \citep{2022MNRAS.517.5413H}, can result in a less prominent cosmic ray pressure. We will explore these aspects in future studies.

Finally, we are only using the observations of the hot phase of the galaxy. To further break the degeneracy, we would need observations from other phases of the gas, which will be left for future work.

\subsection{Possible further observational probes and model explorations}
\label{s:observation}
To refine the models presented in this study, a more comprehensive examination of X-ray properties and thorough comparisons with multi-phase observations will be necessary. We defer this aspect to future investigations but briefly touch upon potential fruitful directions. While the various models in this study generally align with the X-ray-inferred radially averaged density, temperature, and entropy profiles, the spatial distribution of these properties may exhibit variations, particularly between regions closer to and farther from the jet axis. These differences could be further elucidated through extensive comparisons of X-ray maps.

Additionally, mapping the kinetic properties (inflow/outflow and turbulent velocities versus polar angle) near the jet can provide additional constraints on the models. The thermal characteristics of lower-temperature gas can vary within approximately 30 kpc between runs employing cosmic-ray and thermal jets with different energy fluxes, leading to disparate predictions for the column densities of various ions in different phases.

We have verified that, in our CR jets running with a halo mass below $10^{14} M_\odot$, the predicted $\sim$GeV gamma-ray luminosity from hadronic interactions is below the current observational upper limits \citep{2016ApJ...819..149A,2019MNRAS.488..280W}.

In contrast, the ``m15-CR7e44'' case experiences an increase in gamma-ray luminosity due to hadronic loss, ranging from $10^{43}$ to $10^{44}$ erg s$^{-1}$ as cooling flows resume. Notably, this range exceeds the  observational constraints. Over 90\% of the gamma-ray flux emanates from the inner 20 kpc, serving as motivation for models proposing the injection of cosmic rays at the jet cocoon shock rather than at the point of jet initiation. This adjustment would lead to a more extensive distribution of cosmic rays, moving away from their concentration at the galactic core (Su et al. in prep.).

\section{Conclusions}\label{sec:conclusions}

In this paper, we have performed a systematic exploration of different AGN jet models that inject energy into massive halos from $10^{12}$ to $10^{15} {\rm M_\odot}$, quenching galaxies and suppressing cooling flows. We considered models with a wide range of jet properties, including: precessing kinetic jets, thermal energy-dominated jets, and cosmic ray jets with fixed energy fluxes. These models were investigated through simulations at a full-halo scale, although they were non-cosmological. The simulations incorporated various physical processes, such as radiative heating and cooling, self-gravity, star formation, stellar feedback from supernovae, stellar mass-loss, and radiation. This comprehensive approach allowed for a dynamic and ``live'' response of star formation and the multi-phase ISM to cooling flows, despite the continuous flux of AGN jets. For more massive halos (m14 and m15), we employed a hierarchical super-Lagrangian refinement scheme, achieving a mass resolution of approximately $\sim 3\times 10^{4}\,{\rm M}_{\odot}$?significantly higher than many preceding global studies.

We summarize our key results in the following point:
\begin{itemize}
\item The jet energy required to quench a halo roughly scales with the free fall energy at the cooling radius, consistent with what we hypothesized from simulations of a single halo mass in \citetalias{2021MNRAS.507..175S}.
With the above scaling, all three jet models (precessing kinetic jets, hot thermal jets, and cosmic ray jets) can quench or significantly suppress the star formation over the surveyed halo mass range from $10^{12}$ to $10^{15} {\rm M_\odot}$ without obviously violating our observational constraints.
\item With a similar energy flux, CR dominant jet always quenches more efficiently than the two other variations due to the extra CR pressure support, which does not radiate away as fast at the cooling radius. When limited to just 10\% of the energy flux scaled following Sc-FF, only the CR dominant jet can quench a $10^{12}{\rm M_\odot}$ halo, consistent with what we see in \citetalias{2021MNRAS.507..175S} for the $10^{14}{\rm M_\odot}$ halo case.
\item Scaling the jet energy flux according to the total cooling rate within the cooling radius may not as accurately yield the correct flux required to suppress the cooling flows. The reason is the total central cooling can be dominated by the ISM gas, which is less relevant to the bulk cooling flows.
\item Lowering the CGM mass or increasing the black hole mass will make ISM cooling even more dominant within the cooling radius. As a result, the two scaling methods (free fall energy flux at the cooling radius and cooling rate within the cooling radius) may yield even more discrepant results.
\item Thermal jets can more effectively regulate the gas on sub-100 pc scale, while the CR jets can more efficiently build up the pressure support around the cooling radius. 

\end{itemize}
To summarize, our work lends support to the notion that quenching, particularly in observed galaxies at $z\sim0$ that are brighter than L* in the galaxy luminosity function, can be achieved within the feasible parameter range of AGN jets. Through this investigation and in conjunction with \citetalias{2021MNRAS.507..175S}, we demonstrate that the effective parameter space, ensuring successful quenching without violating observational constraints, highlights specific jet/cocoon processes and potentially implicates a role for cosmic rays. However, it is essential to acknowledge various caveats (\sref{s:caveat}) that require exploration in future studies, along with more in-depth comparisons with observations (\sref{s:observation}).

\vspace{-0.2cm}
\acknowledgments
KS acknowledge support from the Gordon and Betty Moore Foundation and the John Templeton Foundation via grants to the Black Hole Initiative at Harvard University.  CCH, RSS, and DF were supported by the Simons Foundation through the Flatiron Institute. 
Support for SP and PFH was provided by NSF Research Grants 1911233, 20009234, 2108318, NSF CA-REER grant 1455342, NASA grants 80NSSC18K0562, HST-AR-15800. 
R.E. acknowledges the support from grant numbers 21-atp21-0077 and HST-GO-16173.001-A as well as the Institute for Theory and Computation at the Center for Astrophysics. 
CAFG was supported by NSF through grants AST-2108230, AST-2307327, and CAREER award AST-1652522; by NASA through grants 17-ATP17-0067 and 21-ATP21-0036; by STScI through grant HST-GO-16730.016-A; and by CXO through grant TM2-23005X.
Numerical calculations were run on the Flatiron Institute cluster ``popeye'' and ``rusty'', Frontera with allocation AST22010, and Bridges-2 with Access allocations TG-PHY220027 \& TG-PHY220047. 
This work was carried out as part of the FIRE project and in collaboration with the SMAUG collaboration and the LtU collaboration. SMAUG gratefully acknowledges support from the Center for Computational Astrophysics at the Flatiron Institute, which is supported by the Simons Foundation. LtU collaboration is supported by the Simons Foundation.

\vspace{0.3cm}
\section*{Data Availability statement}
The data supporting the plots within this article are available at reasonable request to the corresponding author. A public version of the GIZMO code is available at \href{http://www.tapir.caltech.edu/~phopkins/Site/GIZMO.html}{\textit{http://www.tapir.caltech.edu/$\sim$phopkins/Site/GIZMO.html}}.

\bibliographystyle{mnras}
\bibliography{mybibs}

\appendix
\normalsize

\label{lastpage}

\end{document}